\documentclass[10pt,twocolumn,twoside]{IEEEtran}
\usepackage{amsmath,amsfonts}
\usepackage{amssymb}
\usepackage{algorithmic}
\usepackage{algorithm}
\usepackage{array}
\usepackage[caption=false,font=normalsize,labelfont=sf,textfont=sf]{subfig}
\usepackage{textcomp}
\usepackage{stfloats}
\usepackage{url}
\usepackage{verbatim}
\usepackage{graphicx}
\usepackage{cite}
\usepackage{framed,multirow}
\usepackage[table,xcdraw]{xcolor}

\begin{document}

\title{Building Flyweight FLIM-based CNNs with Adaptive Decoding for Object Detection}

\author{
 \IEEEauthorblockN{Leonardo~de~Melo~Joao*\IEEEauthorrefmark{1}, Azael~de~Melo~e~Sousa\IEEEauthorrefmark{1}, Bianca Martins dos Santos\IEEEauthorrefmark{2}, Silvio~Jamil~Ferzoli~Guimarães\IEEEauthorrefmark{3}, Ewa Kijak\IEEEauthorrefmark{4}, Jancarlo~Ferreira~Gomes\IEEEauthorrefmark{2}, Alexandre~Xavier~Falc\~{a}o\IEEEauthorrefmark{1}}
 
 \IEEEauthorblockA{\IEEEauthorrefmark{1}Institute of Computing, State University of Campinas, Campinas, 13083-872, São Paulo, Brazil}
 
 \IEEEauthorblockA{\IEEEauthorrefmark{2}School of Medical Sciences, State University of Campinas, Campinas, 13083-872, São Paulo, Brazil}
 
 \IEEEauthorblockA{\IEEEauthorrefmark{3}Institute of Computing Sciences, Pontifical Catholic University of Minas Gerais, Belo Horizonte, 30535-901, Minas Gerais, Brazil}

 \IEEEauthorblockA{\IEEEauthorrefmark{4}University of Rennes, IRISA, Inria, 35000 Rennes, France\\\{l228118, a180784, b120382\}@dac.unicamp.br\\sjamil@pucminas.br ewa.kijak@irisa.fr \\ \{jgomes, afalcao\}@unicamp.br}
}

\maketitle
\begin{center} \bfseries EDICS Category: TEC-MLI \end{center}

\begin{abstract}
State-of-the-art (SOTA) object detection methods have succeeded in several applications at the price of relying on heavyweight neural networks, which makes them inefficient and inviable for many applications with computational resource constraints. This work presents a method to build a Convolutional Neural Network (CNN) layer by layer for object detection from user-drawn markers on discriminative regions of representative images. We address the detection of Schistosomiasis mansoni eggs in microscopy images of fecal samples, and the detection of ships in satellite images as application examples. We could create a flyweight CNN without backpropagation from very few input images. Our method explores a recent methodology, Feature Learning from Image Markers (FLIM), to build convolutional feature extractors (encoders) from marker pixels. We extend FLIM to include a single-layer adaptive decoder, whose weights vary with the input image -- a concept never explored in CNNs. Our CNN weighs thousands of times less than SOTA object detectors, being suitable for CPU execution and showing superior or equivalent performance to three methods in five measures.
\end{abstract}


\section{Introduction}\label{sec:intro}
Most state-of-the-art (SOTA) algorithms for object detection rely on deep neural networks, requiring large annotated datasets to train heavyweight models with millions of parameters that demand powerful computational resources in GPUs~\cite{zaidi2022survey}. Even models based on few-shot learning, which require considerably less training images, are heavyweight~\cite{huang2021survey} and unsuitable for tasks with computational resource constraints or very few annotated images.

Pretraining a deep-learning architecture with large annotated datasets, such as Imagenet~\cite{deng2009imagenet} and MSCoco~\cite{lin2014microsoft}, and specializing it for a given problem creates heavyweight models hard to simplify. Instead, this work builds on top of a recent methodology, \textit{Feature Learning from Image Markers} (FLIM)~\cite{de2020feature}, to create convolutional feature extractors from image markers, usually drawn by the user on discriminative regions of very few training images (e.g., five). In FLIM, the user indicates attention regions for filter learning by drawing scribbles on representative images, and the filters of a sequence of convolutional layers are obtained from patch datasets -- i.e., patches extracted from marker pixels using the previous layer's output and the exact shape of the filters. Those filters can be learned by patch clustering, for instance, without weight initialization and backpropagation (Section~\ref{sec:method}). 

This work proposes a single-layer and unsupervised adaptive decoder, creating a new FLIM-based method to construct a CNN layer by layer for object detection. The decoder provides an object saliency map on top of which the object detection is performed, and its weights are estimated on-the-fly for each input image according to heuristics that model prior knowledge of the image domain -- a concept never explored in CNNs. By using a tool, \textit{FLIM-Builder}\footnote{\url{https://github.com/LIDS-UNICAMP/FLIM-Builder}.}, we explore more active user participation when determining the architecture of the initial layers of a FLIM-based encoder. The user can visually inspect the filter's (kernel's) outputs to eliminate redundancy and irrelevant information with manual kernel selection. Since our decoder is unsupervised, the user can also verify the model's results after each intervention, allowing to evaluate the impact of his/her actions in real-time.

\begin{figure}[t!]
    \centering
    \begin{tabular}{c c}
         \includegraphics[width=0.37\linewidth]{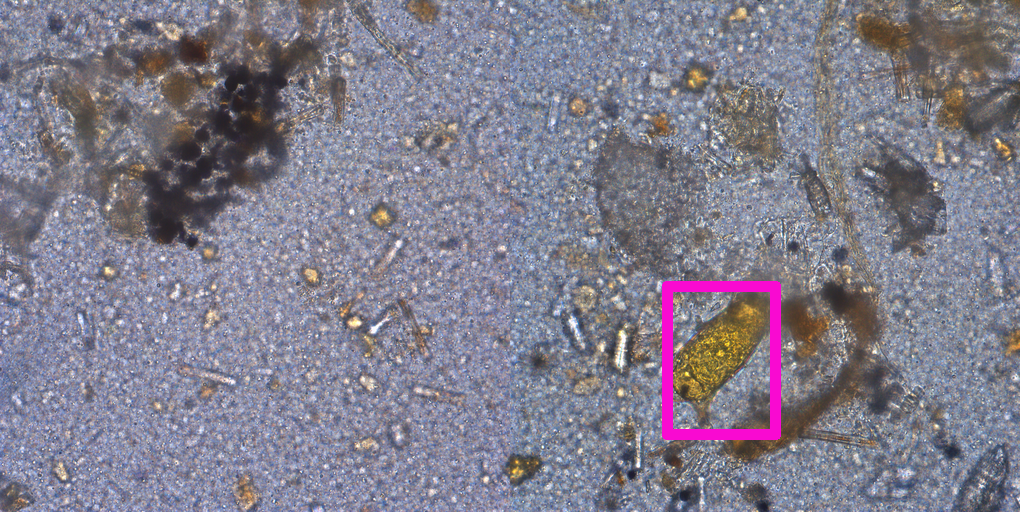} &
         \includegraphics[width=0.37\linewidth]{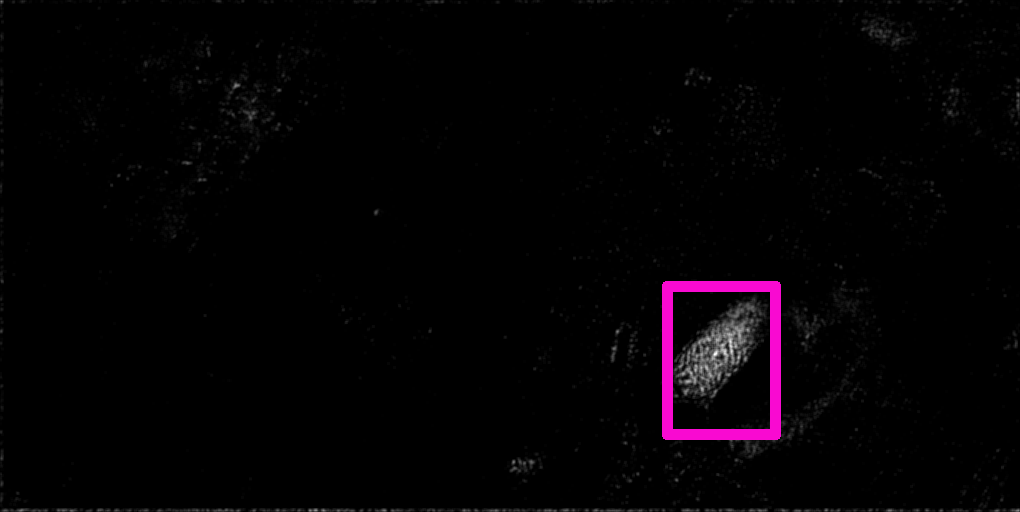} \\
         \includegraphics[width=0.37\linewidth]{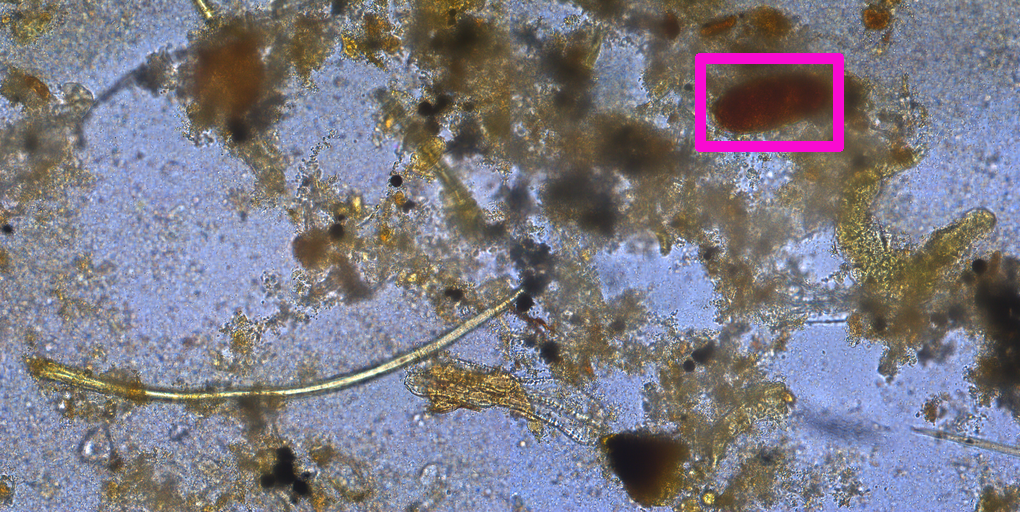} &
         \includegraphics[width=0.37\linewidth]{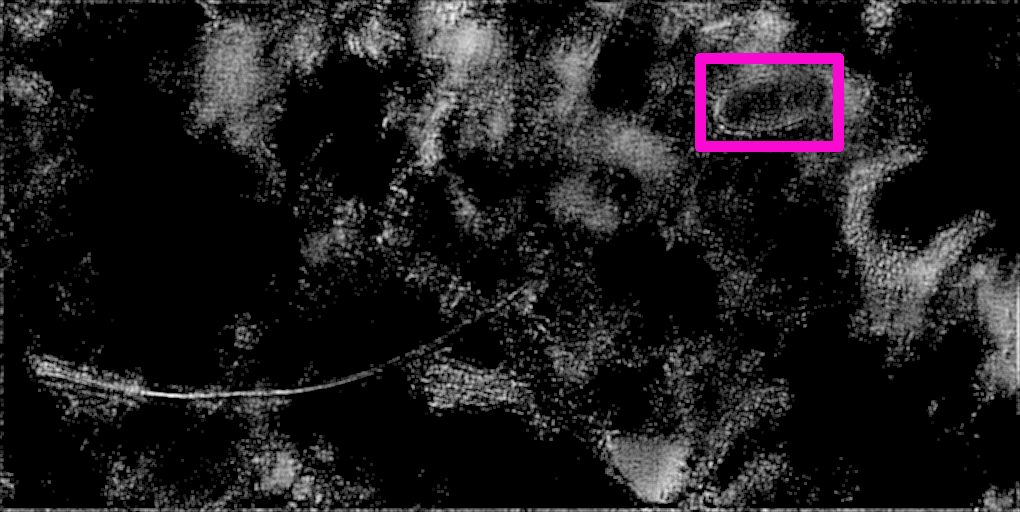} \\
         (a) & (b) \\
     \end{tabular}
    \caption{The activation maps of a  kernel in two training images. (a) Original images; (b) Activation maps for a given kernel. The parasites are highlighted by pink boxes. In the proposed adaptive decoder, the same kernel is considered  foreground dominant for the top image, and background dominant for the bottom one.}
    \label{fig:foreground-background-kernel}
\end{figure}

For an encoder's layer under construction, our decoder weights the selected kernels as foreground or background dominant according to their mean activation and a threshold. This characteristic makes it adaptive to each input image (Figure \ref{fig:foreground-background-kernel}). The decoder performs a point-wise convolution with the adaptive weights, adding activation maps from foreground kernels and subtracting activation maps from background kernels to output a single object saliency map. The objects are detected by thresholding the map and filtering components by area. A minimum bounding box containing each component describes its position. It is important to highlight that the unsupervised and adaptive decoder is essential for enabling the kernel selection strategy (Section~\ref{sec:method}). The user can evaluate the results for the selected training images and decide to add or not another layer, whose kernels are estimated by the FLIM algorithm and selected by the user based on the visual analysis of the saliency map. By that, the user can determine the depth of the CNN architecture for the given problem. 

Although our flyweight CNN is more closely related to saliency estimators than to object detectors, we compare it with state-of-the-art (SOTA) methods from both areas: \textit{$U^2Net$}~\cite{qin2020u2}, which uses nested U-Nets to capture object saliency in multiple scales; SelfReformer~\cite{yun2022selfreformer}, which uses a transformer backbone and a patch-based decoder with a global-context module to create high-resolution saliency maps; and  \textit{DETReg}~\cite{bar2021detreg}, a few-shot-learning object detector implemented as an extension of the Detection Transformer~\cite{carion2020end}. Such methods are heavyweight networks, pretrained with thousands of images and fine-tuned in the same training set used to build our CNN.

For validation, we address the detection of \textit{Schistosoma mansoni}'s eggs in microscopy images~\cite{santos2019tf} and of ships in satellite images~\cite{ships2018}. The first one motivates the construction of tiny models, since a single exam produces thousands of images with millions of pixels each that must be processed within a few minutes in a non-expensive and ultimately embedded machine. The second application illustrates here that the method is suitable for other object detection problems.

For these problems, we could create a flyweight CNN from scratch for object detection using only five training images with user-drawn scribbles -- i.e., with no pixel-wise annotation and no backpropagation. Our model is thousands of times smaller than lightweight models, being efficiently executed on CPU, with only two executions in parallel using our over-the-counter computers being able to fulfill the speed requirement of the laboratory routine for parasite egg detection. 

For the ship detection dataset, our approach achieved competitive results with the three SOTA methods using five metrics. For the in-house parasite dataset (whose images were obtained by processing fecal samples with the \textit{TF-Test Quantified} technique~\cite{santos2019tf}), our approach outperformed the SOTA methods in all five metrics. Additionally, we show kernel selection's importance % and each added layer's impact 
in our ablation studies.

As main contributions, we present: (i) a strategy to build a FLIM encoder layer by layer with kernel selection; (ii) the first unsupervised and adaptive decoder for CNNs; and (iii) a novel flyweight CNN for the detection of \textit{S. mansoni}'s eggs that dismisses GPU execution and outperforms large deep models. 

\section{Related Work}\label{sec:related-work}
     Deep-learning-based object detectors often use a backbone pretrained on ImageNet and fine-tune the model with training samples of the given application~\cite{liu2020deep,huang2021survey}. Few-shot-learning-based approaches additionally pretrain the model in large object-detection datasets before fine-tuning it with a few samples for the given application~\cite{kang2019few,wu2020multi,yan2019meta}. In both cases, the backbone is pretrained on one or more large annotated datasets using a loss function suitable for classification. Alternatively, current object detectors create weak labels to adopt a self-supervised approach for object detection~\cite{bar2021detreg,o2020unsupervised,wei2021aligning,yang2021instance}. They execute an unsupervised object detector on ImageNet to create weak labels and then pretrain the model on ImageNet using the weak labels and a detection loss. Among these self-supervised approaches, the Detection with Transformer using Region priors (DETReg)~\cite{bar2021detreg} has achieved the best results on MSCOCO \cite{lin2014microsoft} -- the most popular dataset for object detection. DETReg uses Selective Search ~\cite{uijlings2013selective} to create weak labels and adopts a class-agnostic model (i.e., estimates bounding boxes around objects without class knowledge). The authors adapt the model for the expected classes by incorporating a classification head. Two problems with this method are the model's size, which has around 40 million parameters and requires expensive GPU execution, and the fixed number of bounding boxes per image, which results in many false positives --  higher the number of false positives, higher is the processing time for object identification. Even though non-maximum suppression may amend the second problem, the model's size makes it unsuitable for certain applications, such as the detection of parasite eggs in the laboratory routine at a reasonable cost for developing countries.

   Class-agnostic approaches can also be implemented by combining a Salient Object Detector (SOD) with an object classifier. SOD methods highlight objects that stand out given observer-defined salient features. The SOTA approaches are based on deep learning~\cite{wang2021salient}. They usually adopt a backbone pretrained on ImageNet, and then the model is trained on a large annotated dataset -- DUTS~\cite{wang2017learning} is the most popular with ten thousand images. Among the recent approaches, U²Net~\cite{qin2020u2} achieved outstanding performance when considering a reduced number of parameters and small number of training images. The method proposes multiple U-shaped blocks to explore multiple scales. By reducing the image size at every step, the number of parameters is kept low. We also evaluated the pre-published SelfReformer (SR)~\cite{yun2022selfreformer}, which has achieved the highest scores in most SOD benchmarks. SelfReformer uses the Pyramid Vision Transformer~\cite{wang2021pyramid} as backbone for improving long-range information dependency, uses Pixel Shuffle~\cite{shi2016real} instead of pooling for keeping fine segmentation details, and frames saliency detection in a patch-wise manner, which relies on a global-context branch to feed information to the local-context patch-based branch. However, U²Net and SR require expensive GPU for parasite egg detection in a viable time. We solve this problem using the FLIM methodology~\cite{de2020feature}, being the first FLIM-based method for object detection.

\section{FLIM-based CNNs with adaptive decoding}\label{sec:method}

\begin{figure*}[t!]
    \centering
         \includegraphics[width=0.8\textwidth]{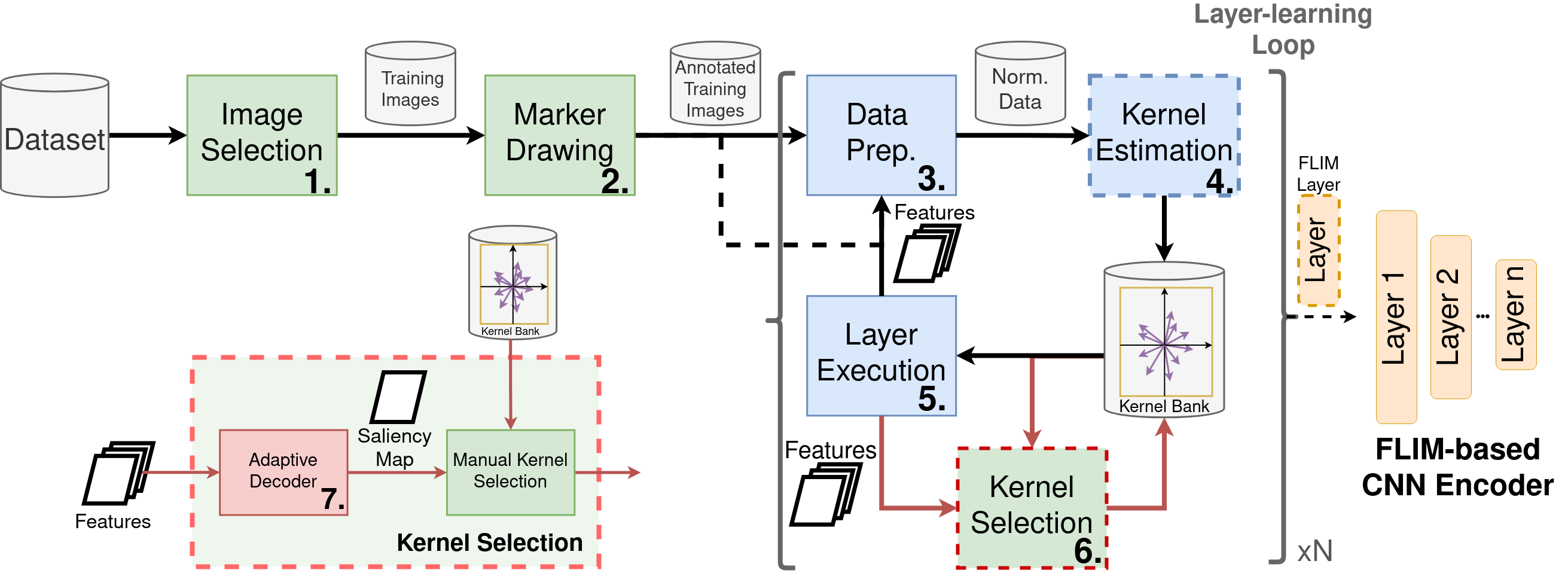} 
     \caption{Diagram depicting the steps to create a flyweight CNN with a FLIM-based encoder and an unsupervised and adaptive decoder. Green boxes indicate user interaction in the step. For a regular FLIM encoder learning, the loop depicted by the red arrows would be skipped.}
    \label{fig:training-diagram}
\end{figure*}

FLIM is a methodology to create feature extractors as a sequence of convolutional layers by estimating kernels from patch datasets defined from image markers at each layer's input (Figure~\ref{fig:training-diagram}). The patches are centered at marker pixels using the previous layer's output features. Each layer aims to separate different object/class properties into distinct regions along the activation channels. The effectiveness of such an operation can be revealed by a point-wise convolution at the output of any layer, decoding the activation channels into an object saliency map. We propose an adaptive decoder, allowing the user to select kernels from the estimated ones, and use the adaptive decoder to evaluate the selected kernel bank at the output of each layer. For object detection, the CNN consists of a FLIM-based encoder followed by a final instance of the adaptive decoder. Each step is explained next together with formalization, insights, and details about the adaptive decoder.
    
\begin{enumerate}
  \item \textbf{Image selection} - Given a training image set of a target problem, we assume redundancy allows us to select a few representative images for encoder training. This step may involve data visualization to guide manual image selection by the expert. For this work, the designer manually selects a small number of representative images (e.g., five images) -- \textit{i.e.}, images with a large range of intra-class variability among objects of the target class, so the class is well represented (e.g., Figure \ref{fig:training-samples} illustrates parasites highly distinct in texture and color). In the training set, at least one example of each distinct characteristic is expected to be present.

\begin{figure}
    \centering
    \subfloat[]{\includegraphics[width=0.32\linewidth]{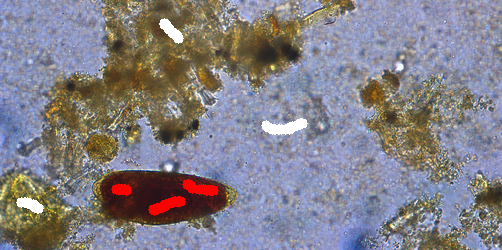}}
    \hfill
    \subfloat[]{\includegraphics[width=0.32\linewidth]{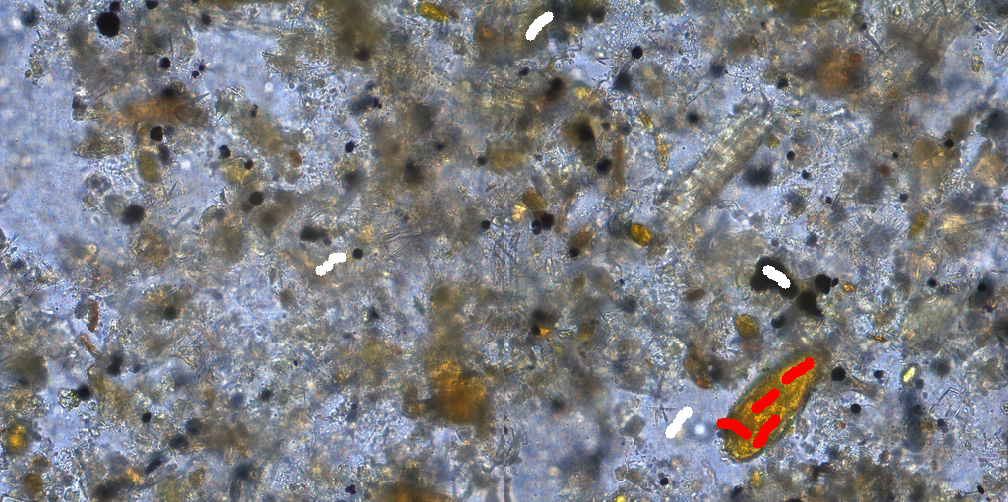}}
    \hfill
    \subfloat[]{\includegraphics[width=0.32\linewidth]{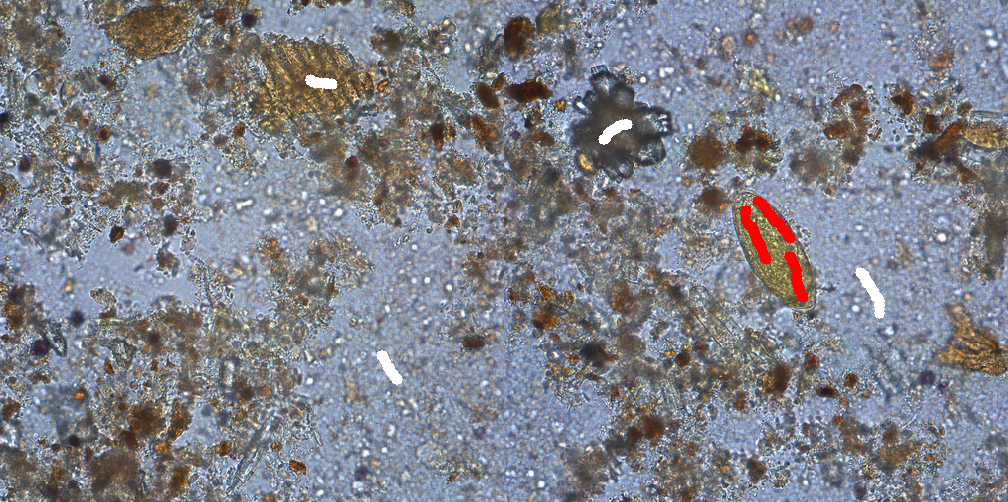}}
    \hfill
   \caption{User-drawn scribbles inside and outside parasite eggs with distinct characteristics. Although we do not use class information in the scribbles, the figure shows red scribbles inside objects and white scribbles in the background. The training set must have representative examples of all intra-class variability for suitable feature extraction design.}
   \label{fig:training-samples}
\end{figure}
  
  \item \textbf{Marker drawing} - In the representative images, we assume there are discriminating regions that represent each class. These relevant regions can be indicated by markers, usually user-drawn scribbles. There is yet to be a formalization of annotation strategies and their impact. All proposed FLIM-based methods have used free-drawn scribbles as image markers, but as presented in~\cite{cerqueira2023building}, having largely unbalanced marker sizes can hamper the model's performance. In this work, the designer freely hand draws scribbles in all object instances and distinct background regions. The markers drawn in Figure~\ref{fig:training-samples} are the ones used in our experiments.

  \item \textbf{Data preparation} - Data preparation requires marker-based normalization (Section~\ref{subsec:flim-intuitions}) and marker scaling onto each new layer's dimension. In this work, we do not use strides, preserving the image scale in all layers. 
  
  \item \textbf{Kernel Estimation} - Given the architecture of a convolutional layer, and the layer's input (original image or activations from the previous layer), patches centered at the marker pixels are extracted according to the desired kernels' shape for that layer. The resulting patch dataset contains the candidate kernels. We estimate kernels as cluster centers of a k-means on the patch dataset, obtaining a given number of kernels per marker $k_m$ and so avoiding unbalanced kernel sets per marker.  Then, the total number of markers is reduced using another K-means (or Principal Component Analysis~\cite{sousa2021cnn}) to fit the number of filters $k_l$ required in the defined architecture for a given layer. Such a strategy has been used in other works~\cite{cerqueira2023building}. One may choose a given number of kernels per class~\cite{de2020learning} but most works do not consider class information.

  \item \textbf{Layer execution} - The convolutional layer is executed to obtain new image features. As a novelty in the pipeline, the adaptive decoder can be used at the output of any layer for kernel selection, as explained. If the user identifies an abundance of false positives in the saliency maps of a current layer, a new convolutional layer is added to the encoder.
  
  \item \textbf{Kernel selection} - This work uses the proposed adaptive decoder to create an object saliency map from any training image at the layer's output. The network designer can visualize the activation maps for each of the $k_l$ estimated kernel and select which kernels to keep in that layer (Figure \ref{fig:kernel-selection-example-b}). Kernel selection affects the saliency map, allowing the visual evaluation of the overall quality of the selected kernel set (Figure \ref{fig:kernel-selection-example-c}). Often, the user goes through the activation thumbnails and closely inspects the kernel activation maps with more distinct characteristics. The resulting kernel bank is then used to create the input for designing the next layer.

In parasite egg detection, for example, two observations can guide kernel selection: (i) foreground kernels that together activate all eggs in the training images and (ii) pairs of foreground and background kernels that activate the same impurities. Since a foreground kernel may not activate all eggs (Figure \ref{fig:foreground-background-kernel}), multiple kernels are required to satisfy (i). Observation (ii) implies that both kernel types should be selected to reduce the number of false positives (Figure \ref{fig:complementary-background}) since the decoder subtracts background and adds foreground activation maps to create the object saliency map. If the result is not satisfactory, the user can change the selection.  

\begin{figure}
    \centering
    \subfloat[]{\includegraphics[width=0.32\linewidth]{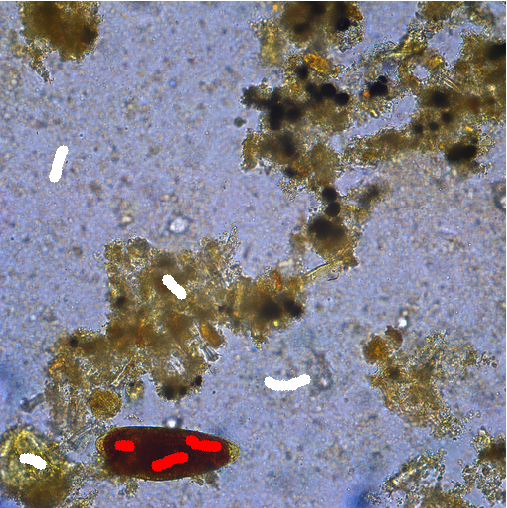}}
    \hfill
    \subfloat[]{\includegraphics[width=0.29\linewidth]{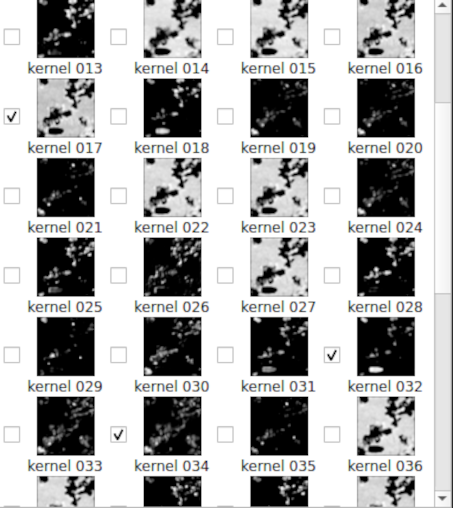}
    \label{fig:kernel-selection-example-b}}
    \hfill
    \subfloat[]{\includegraphics[width=0.32\linewidth]{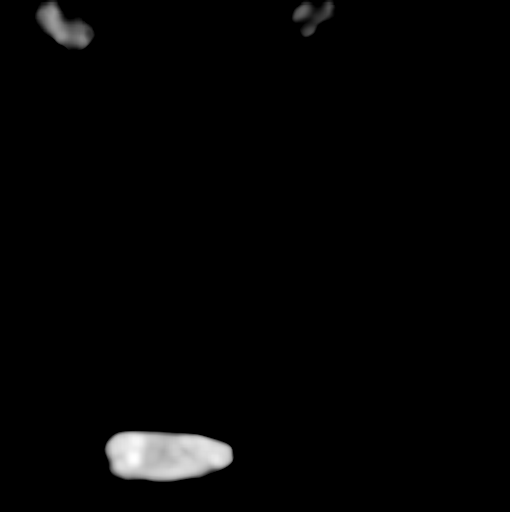}
    \label{fig:kernel-selection-example-c}}
    \hfill
   \caption{Example of the informations the user interacts to when selecting kernels. (a) Original image with markers; (b) List of kernels from the third layer, with a few of them selected (check-boxes); (c) Example of a decoded salient map considering the user's selection. The individual activations can also be viewed full-size by the user.}
   \label{fig:kernel-selection-example}
\end{figure}

  \item \textbf{Adaptive decoder} - The features (activation maps) are combined into an object saliency map using the adaptive weights of a point-wise convolution followed by ReLU activation. Such weights are positive for foreground kernels and negative for background ones. However, the kernel's sign may change for each image as automatically determined according to the application. The details are presented in Section \ref{subsec:flim-decoder}.
  
\end{enumerate}
\subsection{Formalization and Insights}\label{subsec:flim-intuitions}

\textbf{Images, adjacency relations, and image patches:} Let $\mathbf{X} \in \mathbb{R}^{h\times w\times c}$ be an image with $h \times w$ pixels and $c$ channels. A pixel at position $(i,j)$, $i \in \{1, h\}, j \in \{1,w\}$, can be represented by its feature vector $\mathbf{x}_{ij} \in \mathbb{R}^c$, and $x_{ijb} \in \mathbb{R}, b \in \{1, c\}$, denotes the b-th channel (feature) value for pixel (i,j). 
A patch $\mathbf{p}_{ij} \in \mathbb{R}^{k\times k\times c}$, is a sub-image composed of the $c$ features of all $k\times k$ pixels adjacent to the one in position $(i,j)$. 

\textbf{Filters, convolutions and mathematical interpretations:} A kernel (or filter) $\mathbf{k} \in \mathbb{R}^{k\times k\times c}$ is a "moving subimage" with the same shape than a patch. The convolution\footnote{The original concept is simplified here to preserve the image domain and assume the adjacency relation is already reflected in each axis.} of an image with a filter can be described as $\mathbf{Y} = \mathbf{X} \star \mathbf{k}$. Assuming zero padding and no stride, $\mathbf{Y} \in \mathbb{R}^{h\times w\times 1}$ and $y_{ij} \in \mathbf{Y}$.  

Let $\tilde{\mathbf{p}}_{ij} \in \mathbb{R}^d$ (resp. $\tilde{\mathbf{k}} \in \mathbb{R}^d$), with $d=k \cdot k \cdot c$ be the flattened representation of $\mathbf{p}_{ij}$ (resp. $\mathbf{k}$). The convolution result $y_{ij}$ can then be seen as the dot product between the vectors representing the filter and the patch centered at the pixel (i,j):

\begin{eqnarray}\label{eq:pixelconv}
y_{ij} & = & \langle \tilde{\mathbf{p}}_{ij}, \tilde{\mathbf{k}}\rangle \\
\mbox{with} \;\; \langle \tilde{\mathbf{p}}_{ij}, \tilde{\mathbf{k}}\rangle & = & \|\tilde{\mathbf{p}}_{ij}\| \| \tilde{\mathbf{k}}\| \cos{\theta}, \nonumber
\end{eqnarray}
where $\theta$ is the angle between both vectors. Equation~(\ref{eq:pixelconv}) has an important geometric interpretation when filtering images for feature extraction. Filter $\mathbf{\tilde{k}}$ can be interpreted as a vector orthogonal to a hyperplane passing through the origin of $\mathbb{R}^{d}$ and $\tilde{\mathbf{p}}_{ij}$ is a point in $\mathbb{R}^{d}$. When the norm $\|\tilde{\mathbf{k}}\|=1$, $y_{ij}$ is the distance between $\tilde{\mathbf{p}}_{ij}$ and the hyperplane of $\mathbf{\tilde{k}}$ --- it may be positive or negative depending on which side of the hyperplane the point $\tilde{\mathbf{p}}_{ij}$ is. When $\tilde{\mathbf{k}}$ does not have unit norm, $y_{ij}$ is amplified by its magnitude. The convolution between an image and a filter slides the filter over the image and computes the signed angular distance $y_{ij}$ for every pixel. When $\mathbf{X}$ is filtered by a set (bank) with $m$ kernels, the resulting image $\mathbf{Y} \in \mathbb{R}^{h\times w\times m}$ has multiple channels, each channel being the result of filtering by one kernel. 

Another insightful interpretation appears when the filtered image $\mathbf{Y} \in \mathbb{R}^{h\times w\times m}$ is further transformed by an \textbf{activation function} $\phi$ defining an image $\mathbf{A} \in \mathbb{R}^{h\times w\times m}$, where $\mathbf{a}_{ij}=\phi(\mathbf{y}_{ij})$. A commonly used example is the Rectified Linear Unit (ReLU) operation, such that $\mathbf{a}_{ij} = \max\{\mathbf{0},\mathbf{y}_{ij}\}$.

We may interpret the activation $\mathbf{a}_{ij}$ as a similarity between the patch $\mathbf{p}_{ij}$ and the local visual pattern represented by $\mathbf{k}$, which motivates kernel estimation from marker pixels drawn on discriminative regions.

By clustering patches $\tilde{\mathbf{p}}_{i'j'}$ from marker pixels $(i',j')$ and defining kernels $\tilde{\mathbf{k}}$ as cluster centers, Equation~(\ref{eq:pixelconv}) should create positive activations for patches similar to the kernel that represents their cluster. This requires a bias to place the hyperplane of $\tilde{\mathbf{k}}$ at the right location. However, we may eliminate the need for bias computation by applying marker-based normalization before convolution and activation. This operation is explained next.

\textbf{Marker-based Normalization:} By drawing strokes on discriminative regions of a few representative images to a given problem, one can use their pixel values for image normalization before convolution and activation. For feature extraction, markers should indicate local visual patterns that discriminate among image categories/objects of interest.

\begin{figure}
    \centering
    \subfloat[]{\includegraphics[width=0.8\linewidth]{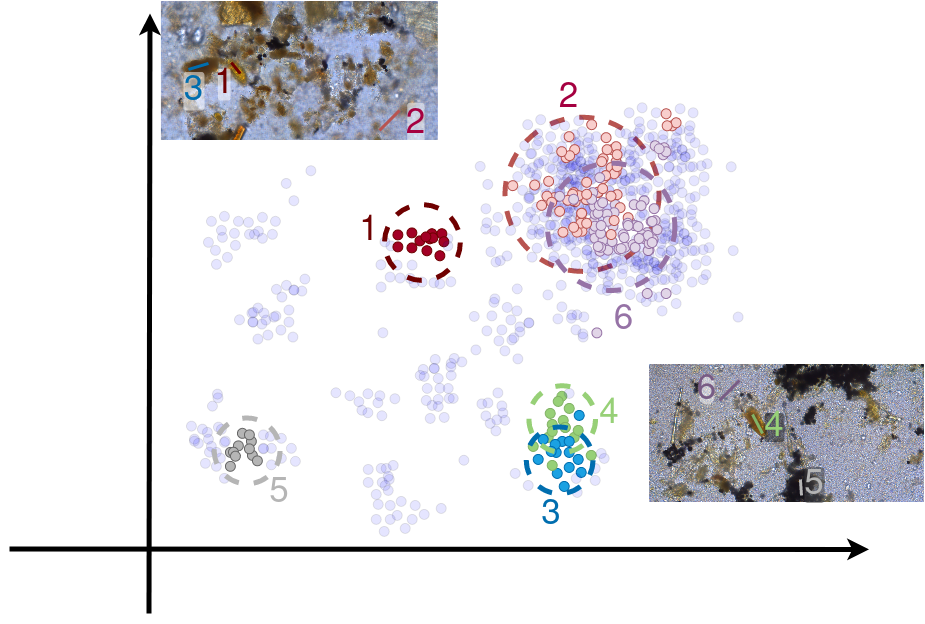}}
    \hfill
    \\
    \subfloat[]{\includegraphics[width=0.48\linewidth]{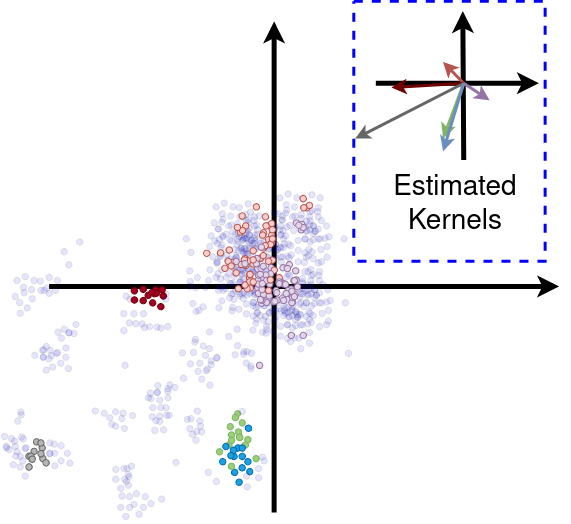}}
    \hfill
    \subfloat[]{\includegraphics[width=0.48\linewidth]{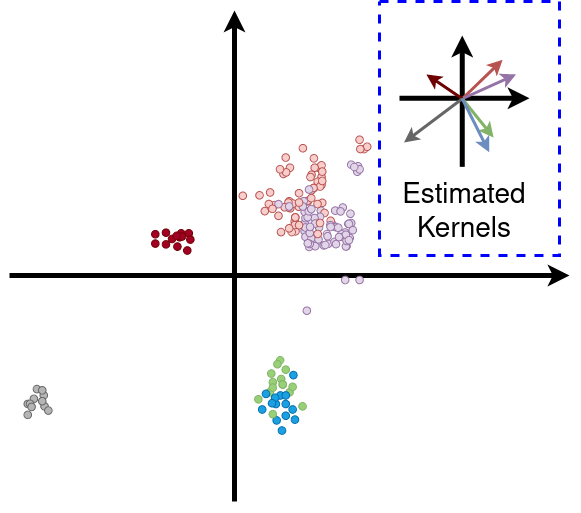}}
    \hfill
   \caption{A 2D feature space illustrating the effect of marker-based undersampling and normalization. Each data point represents a patch centered at a pixel. (a) Data distribution undersampled by image markers, (b) Normalized feature space using all pixels and (c) using marker pixels only. Cosine distances between kernels estimated from the cluster centers in (c) are higher than those from (b), increasing the discriminative power of the operation.}
   \label{fig:markers-featurespace}
\end{figure}

Let $\mathcal{X}$ be a small set of images annotated by markers, $\mathbf{X} \in \mathcal{X}$ be a training image, $\mathcal{M}(\mathbf{X})$ be the set of marker pixels drawn on image $\mathbf{X}$, and $\mathcal{M}$ be the union $\bigcup\limits_{\mathbf{X}\in \mathcal{X}} \mathcal{M}(\mathbf{X})$ of all marker sets. We can transform any image $\mathbf{X}$ into a normalized image $\mathbf{\hat{X}} \in \mathbb{R}^{h\times w\times m}$, where
\begin{align}
\label{eq:mb-normalization}
  \hat{x}_{ijb} & = \frac{x_{ijb} - \mu_b}{\sigma_b + \epsilon},\\
  \notag
  \mu_b  & = \frac{1}{|\mathcal{M}|} \sum_{\substack{i,j \\ x_{ijb}\in \mathcal{M}(\mathbf{X})}} x_{ijb}, \\
  \notag
  \sigma^2_b & = \frac{1}{|\mathcal{M}|} \sum_{\substack{i,j \\ x_{ijb} \in \mathcal{M}(\mathbf{X})}} \left(x_{ijb}-\mu_b\right)^2,
\end{align}
for a small $\epsilon>0$. Marker-based normalization is equivalent to Z-score normalization from marker pixels rather than all training image pixels. Data normalization distributes samples (patches centered at pixels) around the origin of the feature space, increasing the cosine distances between groups of similar samples. Since we intend to identify such groups from marker pixels and use their centers for kernel estimation, convolution using such kernels can better activate pixels from distinct objects/classes in different regions of the activation channels, increasing the discriminative power of the operation -- an effect that could not be achieved with regular normalization due to sample unbalancing (Figure \ref{fig:markers-featurespace}).

\textbf{Pooling:} After marker-based normalization, convolution with estimated kernels and activation, pooling can aggregate nearby activation values of patterns related to a same object/category. In this work, we use max-pooling and average-pooling. Assuming the convolution uses a filter bank with $m$ filters, after activation, image $\mathbf{A}$ will have $m$ channels (i.e. $\mathbf{a}_{ij}\in \mathbb{R}^{m}$). 
Max-pooling defines a new image $\mathbf{M} \in \mathbb{R}^{h'\times w'\times m}$, with $m_{ijb} = \max\limits_{x,y = 0 \cdots s} \{a_{(i+x)(j+y)b}\}$ for a pooling window of size $s\times s$ around pixel $(i,j)$. Similarly, for average-pooling, $m_{ijb} = \frac{1}{s\times s}\sum\limits_{x,y = 0 \cdots s} a_{(i+x)(j+y)b}$.
For pooling operations without stride, $h' = h, w' = w$. While pooling is commonly used to reduce the spatial domain in a convolutional layer, using a \emph{stride} factor per spatial direction (distance between pixels in each direction), we do not use stride in this work so the original image resolution is preserved, considering our solution applies a very simple one layer decoder. Therefore, max-pooling can be seen as a morphological dilation, and average-pooling as an average filter.

\textbf{Convolutional layer:} A convolutional layer may have one or more convolutions with filter banks followed by activation, skip connections, residual operations, batch normalization, or any other type of operation commonly used in CNNs.
To keep the model simple, we use the following sequence of operations: (1) marker-based normalization, (2) convolution with one filter bank, (3) ReLU activation, and (4) pooling.  

\subsection{Single-layer, unsupervised, and adaptive decoder}\label{subsec:flim-decoder}

\begin{figure}[t!]
    \centering
    \begin{tabular}{c c c}
         \includegraphics[width=0.25\linewidth]{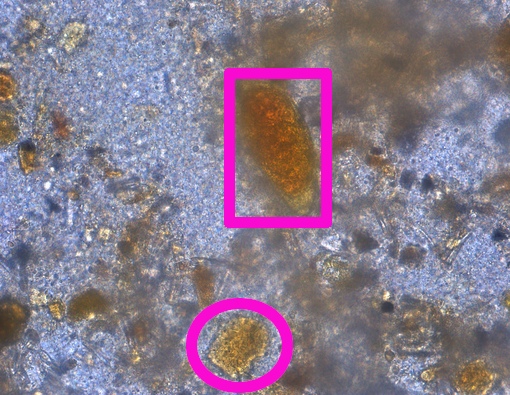} &
         \includegraphics[width=0.25\linewidth]{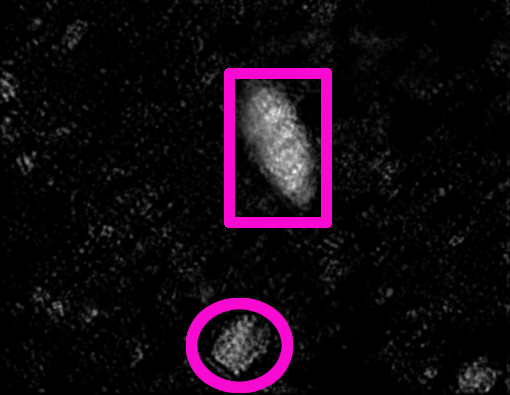} &
         \includegraphics[width=0.25\linewidth]{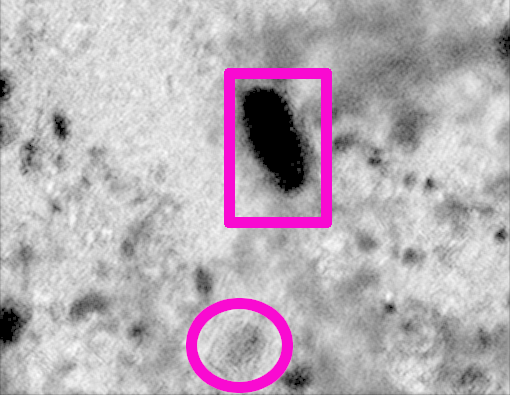} \\
         (a) & (b) & (c)\\
     \end{tabular}
    \caption{Activation maps of a pair of foreground and background kernels for an image. (a) Original image; (b) Foreground kernel activation; (c) Background kernel activation. The egg (in the box) is activated in (b) and is not activated in (c). However, the impurity (in the ellipse) is activated in both and can be eliminated by subtracting (c) from (b).}
    \label{fig:complementary-background}
\end{figure}

Let $\mathbf{A} \in \mathbb{R}^{h\times w\times m}$ be the output of an encoder's layer. A point-wise convolution is a weighted sum of the $m$ channels (activation maps) in $\mathbf{A}$. A point-wise filter is defined by a weight vector $\boldsymbol{\alpha} = [\alpha_1, \alpha_2, ..., \alpha_m] \in \mathbb{R}^{m}$, such that $\alpha_b \in [-1,1]$, $b=1,2,\ldots,m$. The value $|\alpha_b| \in [0,1]$ represents the importance of a channel $b$ (i.e., the importance of the $b$-th kernel that originated that channel) regarding the foreground or background. $\alpha_b$ is positive for foreground activation maps and negative for background. Given that a kernel can activate the foreground for an input image and the background for another one (Figure~\ref{fig:foreground-background-kernel}), we propose adaptation heuristics based on prior domain information to determine when a kernel is foreground- or background-dominant. The decoder subtracts the weighted sum of background-dominant activation maps from the weighted sum of the foreground-dominant ones, followed by ReLU activation (Figure \ref{fig:complementary-background}). 

    Let $H : \alpha_b \to \{-\alpha, 0, \alpha\}$ be a function that maps every weight to a positive, negative, or zero value according to an \emph{adaptation heuristic}. In this work, we present two functions for weight adaptation. Both functions indirectly explore the ratio of the target objects' area over the image's area as prior domain information valid for both applications of interest. Such a ratio is expected to be small. For such, we compute the mean activation values of each $b$-th kernel as $\mu_{\mathbf{A}_b} = \frac{1}{hw} \sum\limits_{a_{ijb} \in \mathbf{A}}{a_{ijb}}$. This function is a simple global average pooling for GPU implementation.

For parasite egg detection, the objects' sizes are consistently and considerably smaller than the sizes of background components. We then adopt a very simple threshold ($\lambda = 0.5$) to distinguish between foreground and background kernels:
    
\[
    \mathbf{H}_p(\mathbf{A}, b)= 
\begin{cases}
    +\alpha,& \text{if } \mu_{\mathbf{A}_b}\leq 0.5\\
    -\alpha, & \text{otherwise.}
\end{cases}
\]

For ship detection, the objects' sizes are inconsistent with ships in different scales and background components with various sizes may appear from some noisy images, asking for a more complex adaptation function. In this case, we compute two thresholds based on the mean activation values $\mu_{\mathbf{A}_b}$ from all channels $b=1,2,\ldots,m$. Let $\boldsymbol{\mu} = \frac{1}{m} \sum_{b=1}^m \mu_{\mathbf{A}_b}$ and $\sigma_{\boldsymbol{\mu}} = \frac{1}{m} \sum_{b=1}^m (\mu_{\mathbf{A}_b}-\boldsymbol{\mu})^2$ be the mean and standard deviation of the channels' mean activation values, respectively. We then define the weights to be:

\[
    \mathbf{H}_s(\mathbf{A}, b)= 
\begin{cases}
    +\alpha,& \text{if } \mu_{\mathbf{A}_b} \leq \bar{\boldsymbol{\mu}} + \sigma_{\boldsymbol{\mu}}\\
    -\alpha, & \text{if } \mu_{\mathbf{A}_b}\geq \bar{\boldsymbol{\mu}} - \sigma_{\boldsymbol{\mu}} \\
    0, & \text{otherwise.}
\end{cases}
\]

\begin{figure}
    \centering
    \subfloat[]{\includegraphics[width=0.22\linewidth]{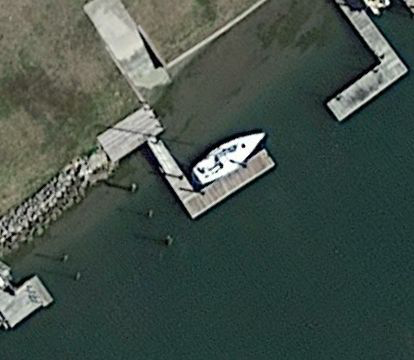}}
    \hfill
    \subfloat[]{\includegraphics[width=0.22\linewidth]{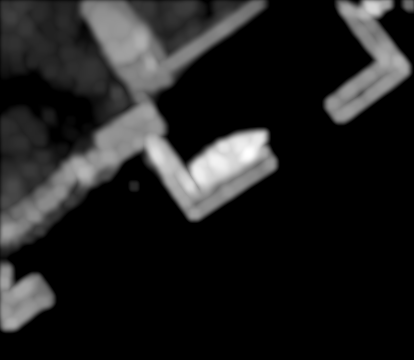}}
    \hfill
    \subfloat[]{\includegraphics[width=0.2\linewidth]{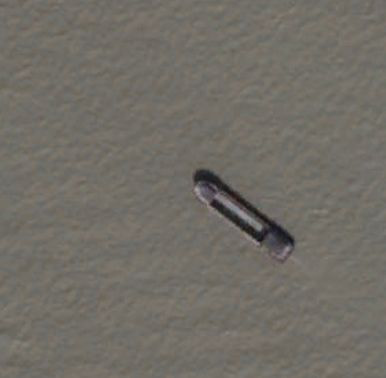}}
    \hfill
    \subfloat[]{\includegraphics[width=0.2\linewidth]{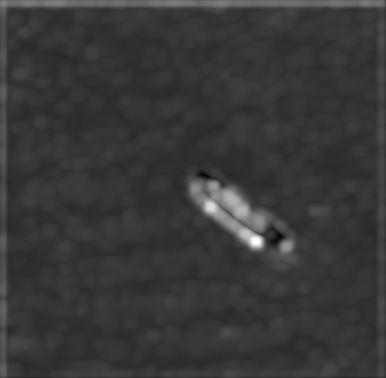}}
    \hfill
    \caption{Cases of neutral kernels. (a,c) Original images; (b,d) Activation Maps.}
    \label{fig:heuristic-neutral}
\end{figure}

By using $\mathbf{H}_s$, we define a range for kernels that are uncertain, \textit{i.e.}, invalid for being foreground and background (Figure \ref{fig:heuristic-neutral}). We also eliminate kernels with no distinctly salient regions, \textit{i.e.}, maps with gray values for most pixels. For that, we consider $\mathbf{H}_s(\mathbf{A}, b) = 0 $ if $0.25 \leq \mu_{\mathbf{A}_b} \leq 0.75$ with standard deviation $\sigma_{\mathbf{A}_b}<0.1$.     

    For this work, we consider $\alpha\in \{-1,1\}$, then $\alpha_b\in \{-1,0,1\}$. The point-wise convolution followed by ReLU activation results in an object saliency map $\mathbf{S}$ defined by:

    \begin{equation}
        \mathbf{S} = \phi(\langle\mathbf{A}, \boldsymbol{\alpha}\rangle)
    \end{equation}
    
    It is worth pointing out that such adaptive point-wise convolutions may be useful in the decoder of other CNN networks. To do so, one must provide an adaptation function (or use the proposed one) and ensure all kernel weights are normalized within $[-1,1]$.
    
\section{Experimental Results}\label{sec:results}
\subsection{Experimental Setup}\label{subsec:params}

    \textbf{\textit{Schistosoma Mansoni} eggs dataset}: The parasite dataset is proprietary and comprises $631$ images containing \textit{S. mansoni} eggs. The images were pixel-wisely annotated by specialists, providing a segmentation mask. The images often have a cluttered background, and fecal impurities sometimes occlude the eggs (Figure \ref{fig:complementary-background}). 
    
    \textbf{Ship Detection Dataset}: The ship dataset~\cite{ships2018} comprises $621$ aerial images containing ships of varying sizes, scales, and colors. Bounding-box ground truth is available on the dataset website, and we manually drew pixel-wise masks to train the models. This dataset contains overlapping ships as different objects in the ground truth, and the saliency-based methods do not handle such cases.

    \textbf{Computer Setup:} Our models were trained and executed on an i7-7700 (CPU), while the heavyweight models used an NVIDIA RTX A6000 (GPU).

\begin{figure}[t!]
    \centering
         \includegraphics[width=0.95\linewidth]{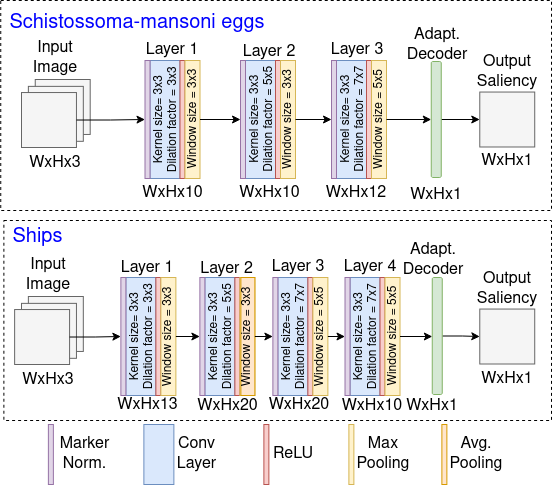} 
    \caption{Resulting CNNs for each dataset: \textbf{Adaptive-}$\mathbf{FLIM_p}$ (top) and \textbf{Adaptive-}$\mathbf{FLIM_s}$ (bottom). Note that the image resolution stays the same due to the absence of stride in pooling and convolution.}
    \label{fig:network-diagram}
\end{figure}

    \textbf{FLIM hyperparameters}: In the kernel estimation step, the number of kernels per marker $k_m$ was fixed to 1 and 5 in all layers of the models for the ship and parasite datasets, respectively. The estimated kernels per layer $k_l$ were $32$ in the first layer and $64$ in the others. After kernel selection, the respective resulting CNNs for each dataset are depicted in Figure~\ref{fig:network-diagram}. Since we are not using stride, we adopted atrous convolutions with varying dilation factors at each layer.
    
    \textbf{Saliency-based bounding boxes:}     
Otsu's thresholding binarizes the saliency maps for extracting bounding boxes around the binary components. For our method, we increased the minimum bounding boxes by $10\%$ for parasite detection. Such adjustment did not improve the other methods and our results in the ship dataset.

    \textbf{Bounding box filtering:} Schistosoma mansoni eggs fall in a particular range of sizes, allowing to discard bounding boxes outside that range. In both datasets, we remove bounding boxes with sizes smaller than one hundred pixels to handle small noise components. The same filtering criteria were applied to all models. Additionally, because DETReg outputs a fixed number (30) of bounding boxes per image, we executed non-maximum suppression (NMS) using an IoU threshold of $0.1$ due to the low chance of overlapping objects.
    
    \textbf{Cross-validation}: We randomly split the datasets into $20$\% for testing and $80$\% for training and validation. A 3-fold cross-validation was adopted for the experiments, using $1$\% (5 images) for training and the rest for validation in each split. The test set was only executed once with the best-performing model considering the validation results of each method in all splits. The training images were selected manually to guarantee a good representation of the real problem.
    
    \textbf{Evaluation metrics}: The methods were evaluated by multiple object detection
metrics related to Intersection over Union (IoU), Precision, and Recall.
The IoU is the ratio between the intersection and union of the
pixels of the predicted bounding boxes and the expected ones.
Let $bb = \{x_1, x_2, y_1, y_2\}$ be a bounding box, such
that $x_2 > x_1, y_2 > y_1$, $B$ be the set of all ground-truth bounding boxes and $bb_p$ be a predicted box. The IoU($bb_p$) can
be defined as: $\mathbf{IoU}(bb_p) = \max\limits_{\forall bb_l \in \mathcal{B}}\{ \frac{| bb_p \cap bb_l |}{| bb_p \cup bb_l|}\}$.

    Bounding boxes with $IoU > \tau$ are considered true positives. Otherwise, they are considered false positives. Let $TP^{\tau}$, $FP^{\tau}$, $FN^{\tau}$ be the total number of true positives, false positives, and false negatives considering $\tau$. Precision is defined as $P^\tau = \frac{TP^\tau}{TP^\tau + FP^\tau}$, and recall as $R^\tau = \frac{TP^\tau}{TP^\tau + FN^\tau}$.
    
For assessing the models, we used the Precision-Recall (PR) curve, considering the mean precision and recall on varying IoU thresholds ranging from $\tau \in [0.50, 0.55, ..., 0.95]$; the mean average precision ($\mu AP$), as proposed for MS Coco considering the same threshold range; average precision ($AP^{\tau'}$); and $F^{\tau'}_2$-score, considering two fixed IoU thresholds $\tau' \in \{0.5,0.75\}$. $AP^{\tau'}$ is the Area Under the Curve (AUC) of the PR-curve up to the fixed threshold, and $\mu AP$ is the mean of all AUCs considering the whole threshold range. $F_\beta\mathrm{-score}^{\tau'} = (1+\beta)^{2} \frac{P^{\tau'}\cdot R^{\tau'}}{\beta^{2} \cdot P^{\tau'}\cdot R^{\tau'}}$ is the harmonic mean of average precision and recall. We used $\beta = 2.0$.
    
\subsection{Ablation on kernel selection and number of layers}

\begin{figure}
    \centering
    \begin{tabular}{c c c}
         \includegraphics[width=0.28\linewidth]{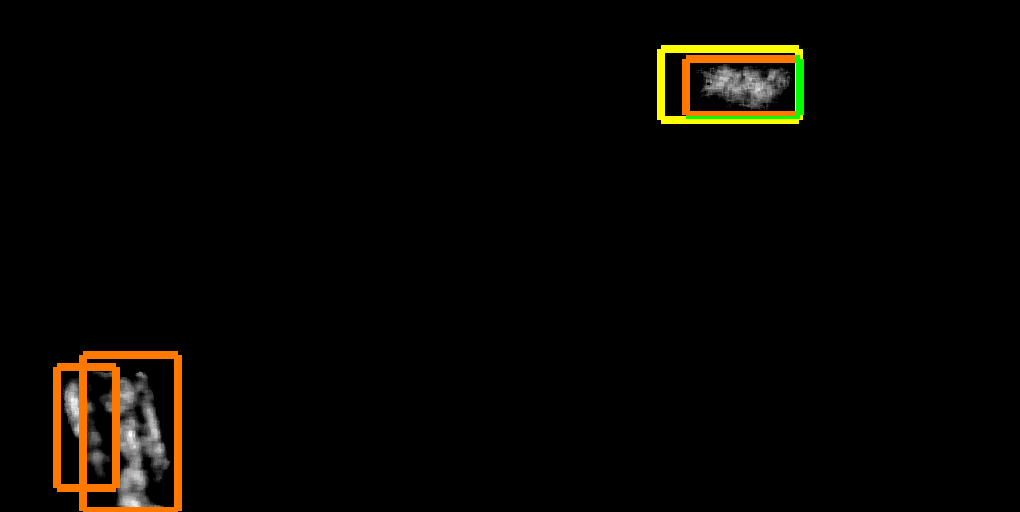} &
         \includegraphics[width=0.28\linewidth]{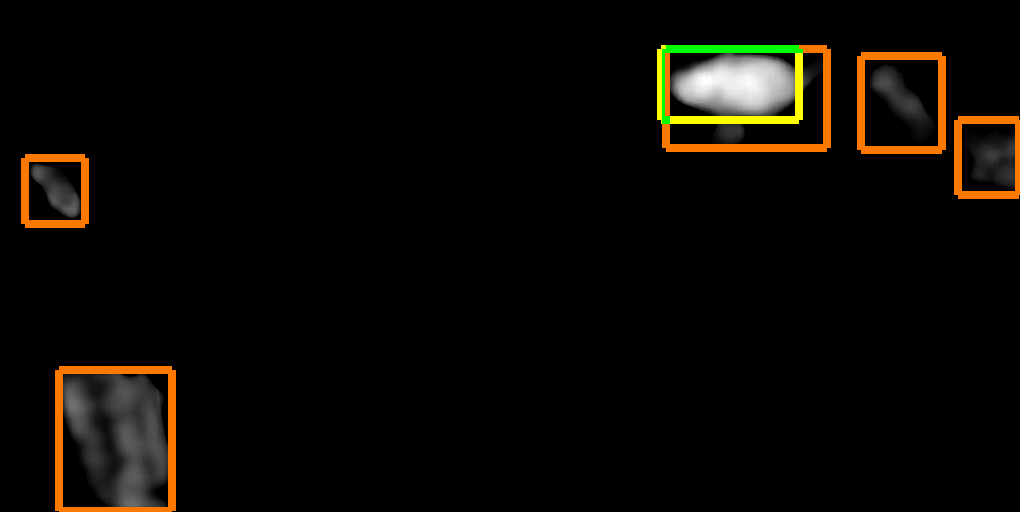} &
         \includegraphics[width=0.28\linewidth]{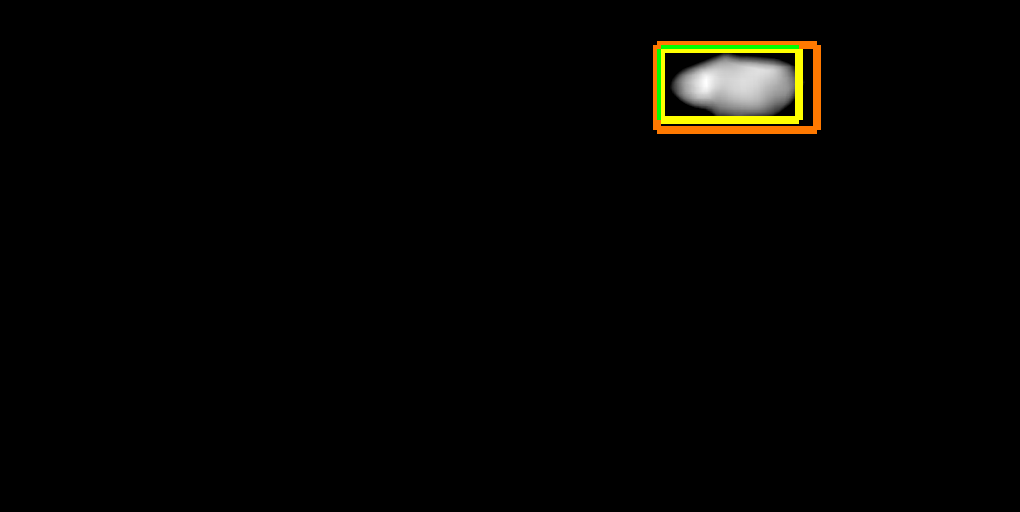} \\
         \includegraphics[width=0.28\linewidth]{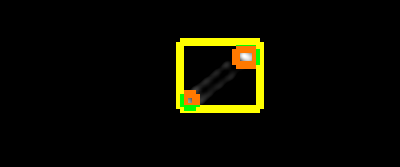} &
         \includegraphics[width=0.28\linewidth]{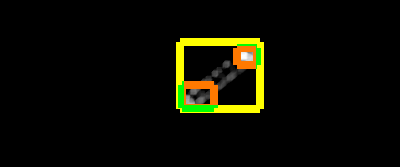} &
         \includegraphics[width=0.28\linewidth]{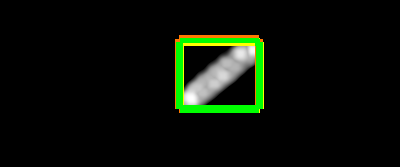} \\
         (a) & (b) & (c)\\
     \end{tabular}
    \caption{Saliency maps using different number of layers for the proposed models. Top: parasite dataset; bottom: ship detection. The ground-truth boxes are in yellow, the predictions in orange, and their intersection in green. (a) First layer, (b) intermediate layer , (c) last layer.}
    \label{fig:quali-ablation-results}
\end{figure}

\textbf{Number of layers:} We evaluated the quality of the saliency maps using our approach for networks with different numbers of layers. Figure \ref{fig:quali-ablation-results} shows a few qualitative examples: Early in the network, for the parasite dataset, objects largely obscured by impurities were often missed, but the number of false positives was low. In the ship dataset, the first layers resulted in maps with more details and heterogeneous saliency, which not fully highlighted objects. 
    
    \textbf{Kernel selection:} 
   To assess the impact of kernel selection in the final model, we executed FLIM with and without manual selection. As presented in Table \ref{tab:ablation-studies}, kernel selection improved the results for both datasets, even though not performing kernel selection still results in models competitive with the state-of-the-art. The improvements were more significant considering $AP^{0.5}$ and $F_{2}^{0.5}$, implying an enhancement in detecting the approximate location of objects rather than an improvement of their spatial extension. Additionally, regarding the number of parameters, the models with kernel selection are more than 10x smaller than the full model for the ships dataset and 20x for the parasite one (Table \ref{tab:model-sizes}), with the full model still being tens of times smaller than the lightweight models, and hundreds or thousands of times smaller than the deep models.
    
    \begin{table}[htb]
 \centering
    \caption{Results of the proposed method with varying number of layers (Adaptive-$FLIM_{1,2,3}$) and the full network with and without~(*) kernel selection.}
    \resizebox{.9\linewidth}{!}{%
    \begin{tabular}{l|c|c|c|c|c}
    \hline
    \rowcolor{gray!25}
    \textcolor{blue}{\textbf{Schistossoma Eggs}}& $F^{0.5}_2 $              & $AP^{0.5}$                & $F^{0.75}_2$              & $AP^{0.75}$               & $\mu$AP \\ \hline 
    Adaptive-$FLIM_1$       &  0.792                    & 0.508                     & 0.568                     & 0.381                     & 0.309  \\ \hline 
    Adaptive-$FLIM_2$       &  \textcolor{green}{0.799} & 0.519                     & \textcolor{green}{0.591}  & 0.401                     & 0.320  \\ \hline 
    Adaptive-$FLIM_p$*        &  0.654                    & \textcolor{green}{0.754}  & 0.470                     & \textcolor{green}{0.440}  & \textcolor{green}{0.419}  \\ \hline
    \textbf{Adaptive-}$\mathbf{FLIM_p}$  &  \textcolor{blue}{\textbf{0.799}}  & \textcolor{blue}{\textbf{0.929}}  & \textcolor{blue}{\textbf{0.630}} & \textcolor{blue}{\textbf{0.488}}   & \textcolor{blue}{\textbf{0.525}}  \\ \hline 
    
    \rowcolor{gray!25}
    \textcolor{blue}{\textbf{Ships}}& $F^{0.5}_2 $      & $AP^{0.5}$                & $F^{0.75}_2$              & $AP^{0.75}$               & $\mu$AP \\ \hline 
    Adaptive-$FLIM_1$       &  0.323                    & \textcolor{green}{0.263}  & 0.205  & 0.089                     & \textcolor{green}{0.139}  \\ \hline 
    Adaptive-$FLIM_2$       &  0.304                    & 0.228                     & 0.199                     & 0.086                     & 0.107  \\ \hline 
    Adaptive-$FLIM_3$       &  \textcolor{green}{0.332} & 0.252                     & \textcolor{blue}{\textbf{0.211}}  & \textcolor{green}{0.092}  & 0.111  \\ \hline
    Adaptive-$FLIM_s$*      &  0.305                    & 0.221                     & 0.178                     & 0.090                     & 0.106  \\ \hline
    \textbf{Adaptive-}$\mathbf{FLIM_s}$  &  \textcolor{blue}{\textbf{0.393}}  & \textcolor{blue}{\textbf{0.322}}  & \textcolor{green}{0.210} & \textcolor{blue}{\textbf{0.145}}   & \textcolor{blue}{\textbf{0.150}}  \\ \hline 
    \end{tabular}
    }
    \label{tab:ablation-studies}
\end{table}
    
\subsection{Quantitative Results}\label{subsec:quantitative}
    Table \ref{tab:quantitative-results-splits} presents the mean and standard deviation of the metrics for all methods using the validation sets of the three splits. \textbf{Adaptive-}$\mathbf{FLIM}$ outperformed all others in all metrics, except for $AP^{0.75}$ in the ship dataset. In this case, U²Net presented the best result. However, U²Net had the highest standard deviation across splits while our solution had the smallest, meaning our method is more consistent and less dependent on the choice of training images.

\begin{figure}
    \centering
    \begin{tabular}{c c}
        \includegraphics[width=0.48\linewidth]{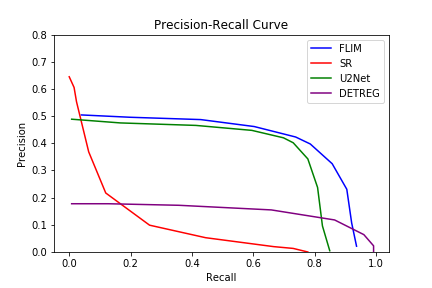} &
        \includegraphics[width=0.48\linewidth]{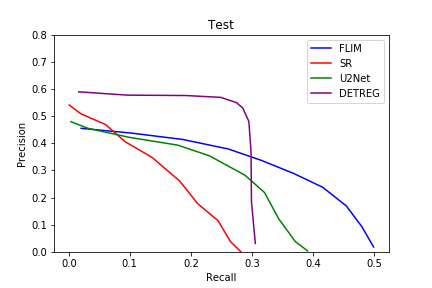} \\
        (a) & (b)
     \end{tabular}
    \caption{Precision-Recall curves for the \textit{Schistossoma} eggs (a) and the ship (b) datasets.}
    \label{fig:PR-curves}
\end{figure}

    The model with the highest $\mu AP$ in the validation set was selected for each method. Table \ref{tab:quantitative-results-test} presents the models' performances on the test set (the unseen images). Our solution had the best results in all metrics for the parasite dataset. DETReg had the lowest values, primarily due to lack of precision (the model outputs many boxes in wrong objects), as seen in the PR curves (Figure \ref{fig:PR-curves}). Self-Reformer had competitive performance when considering a lower \textit{IoU} threshold ($AP^{0.5}, F^{0.5}_2$) but loses performance significantly when the \textit{IoU} requirement is increased; that decrease is also represented in the PR curves, and, as discussed in Section \ref{sec:qualitative-results}, it is primarily due to sub-par segmentation (represented by poorly delineated bounding-boxes). U²Net had the second-best results overall, but more components were missed, hampering their recall values. 
    
    For the ship dataset, all saliency-based solutions obtained poor precision values (mainly affecting the $AP$ metrics), primarily due to small noises being highlighted and their inability to correctly detect multiple connected objects. DETReg can handle cases with connected objects and NMS considerably reduces the number of false positives, however, its recall is considerably lower than the saliency methods. Our method consistently presented either the best or second best result for all metrics, being less precise than DETReg, and having a smaller recall than U²Net for high $IoU$ bounding boxes.
    
    Regarding model size, Table \ref{tab:model-sizes} compiles each network's number of parameters. We also present a size comparison to Lightweight-Network architectures to illustrate the difference. The proposed solution for the parasites dataset is over $17,000$ times smaller than DETReg and almost $40,000$ times smaller than Self-Reformer. Compared to lightweight networks, our approach is $543$ times smaller than SqueezeNet and $1,526$ times smaller than MobileNet. Our solution for the ship dataset is only 3x larger than the one for parasites, still being hundreds of times smaller than the lightweight models.
    
    Regarding execution time, our model executes in around 200ms within a commercial CPU (i7-7700). We did not run extensive time performance comparisons because it would not be suitable for running the large deep models on CPU.
 
\begin{table}[htb]
 \centering
    \caption{Mean and standard deviation of the metrics in the validation set using the three splits. The two best results for each metric are highlighted in blue and green, respectively.}
    \resizebox{.9\linewidth}{!}{%
    \begin{tabular}{l|c|c|c|c|c}
    \hline
    \rowcolor{gray!25}
    \textcolor{blue}{\textbf{Schistossoma Eggs}}& $F^{0.5}_2 $          & $AP^{0.5}$                & $F^{0.75}_2$              & $AP^{0.75}$               & $\mu$AP \\ \hline
    DETReg                                  &  0.562$\pm$ 0.083         & 0.251$\pm$ 0.074          & 0.452$\pm$ 0.053          & 0.183$\pm$ 0.066          & 0.152$\pm$ 0.043  \\ \hline 
    U²Net                                   &  0.682$\pm$ 0.050         & 0.595$\pm$ 0.188          & \textcolor{green}{0.531}$\pm$ 0.034          & \textcolor{green}{0.464}$\pm$ 0.147          & \textcolor{green}{0.325}$\pm$ 0.160 \\ \hline 
    Self-Reformer                           &  \textcolor{green}{0.710}$\pm$ 0.057         & \textcolor{green}{0.655}$\pm$ 0.042          & 0.118$\pm$ 0.031          & 0.021$\pm$ 0.009          & 0.165$\pm$ 0.029 \\ \hline 
    \textbf{Adaptive-}$\mathbf{FLIM_p}$     &  \textcolor{blue}{\textbf{0.792}}$\pm$ 0.001 & \textcolor{blue}{\textbf{0.677}}$\pm$ 0.009& \textcolor{blue}{\textbf{0.609}}$\pm$ 0.027 & \textcolor{blue}{\textbf{0.520}}$\pm$ 0.016 & \textcolor{blue}{\textbf{0.439}}$\pm$ 0.014 \\ \hline 
    \rowcolor{gray!25}
    \textcolor{blue}{\textbf{Ships}}        & $F^{0.5}_2 $              & $AP^{0.5}$                & $F^{0.75}_2$              & $AP^{0.75}$               & $\mu$AP \\ \hline
    DETReg                                  &  0.337$\pm$0.011          & 0.208$\pm$0.010           & \textcolor{green}{0.296}$\pm$0.005           & \textcolor{blue}{0.144$\pm$0.030}& \textcolor{green}{0.127}$\pm$0.012  \\ \hline 
    U²Net                                   &  \textcolor{green}{0.422}$\pm$0.014          & \textcolor{green}{0.290}$\pm$0.009           & 0.235$\pm$0.004           & 0.069$\pm$0.004           & 0.113$\pm$0.006  \\ \hline 
    Self-Reformer                           &  0.192$\pm$0.118          & 0.141$\pm$0.091           & 0.084$\pm$0.059           & 0.024$\pm$0.019           & 0.044$\pm$0.030  \\ \hline 
    \textbf{Adaptive-}$\mathbf{FLIM_s}$     &  \textcolor{blue}{\textbf{0.486}}$\pm$0.007 & \textcolor{blue}{\textbf{0.352}}$\pm$0.011  & \textcolor{blue}{\textbf{0.311}}$\pm$0.024  & \textcolor{green}{\textbf{0.092}}$\pm$0.003  & \textcolor{blue}{\textbf{0.133}}$\pm$0.001  \\ \hline
    \end{tabular}
    }
    \label{tab:quantitative-results-splits}
\end{table}

%_--------------------------------------------------------------------------------------
\begin{table}[htb]
 \centering
    \caption{Test set results considering our proposed solution with kernel selection (\textbf{Adaptive-}$\mathbf{FLIM}$), U²Net, Self-Reformer, and DETReg. The two best results for each metric are highlighted in blue and green, respectively.}
    \resizebox{.9\linewidth}{!}{%
    \begin{tabular}{l|c|c|c|c|c}
    \hline
    \rowcolor{gray!25}
    \textcolor{blue}{\textbf{Schistossoma Eggs}}     & $F^{0.5}_2 $              & $AP^{0.5}$                & $F^{0.75}_2$              & $AP^{0.75}$               & $\mu$AP \\ \hline
    DETReg                                  &  0.634                    & 0.279                     & 0.421                     & 0.146                     & 0.155  \\ \hline 
    U²Net                                   &  0.740                    & 0.531                     & \textcolor{green}{0.609}  & \textcolor{green}{0.405}  & \textcolor{green}{0.335}  \\ \hline 
    Self-Reformer                           &  \textcolor{green}{0.747} & \textcolor{green}{0.688}   & 0.114                     & 0.024                     & 0.227  \\ \hline 
    \textbf{Adaptive-}$\mathbf{FLIM_p}$   &  \textcolor{blue}{\textbf{0.799}}  & \textcolor{blue}{\textbf{0.929}}  & \textcolor{blue}{\textbf{0.630}}   & \textcolor{blue}{\textbf{0.488}}   & \textcolor{blue}{\textbf{0.525}}  \\ \hline 
    \rowcolor{gray!25}
    \textcolor{blue}{\textbf{Ships}}     & $F^{0.5}_2 $                 & $AP^{0.5}$                & $F^{0.75}_2$              & $AP^{0.75}$               & $\mu$AP \\ \hline
    DETReg                                  &  0.261                    & 0.183                     & \textcolor{blue}{0.235}   & 0.130                     & 0.126  \\ \hline 
    U²Net                                   &  \textcolor{green}{0.371} & \textcolor{blue}{0.366}   & 0.164                     & \textcolor{blue}{0.161}   & \textcolor{blue}{0.169}  \\ \hline 
    Self-Reformer                           &  0.251                    & 0.219                     & 0.122                     & 0.105                     & 0.109  \\ \hline 
    \textbf{Adaptive-}$\mathbf{FLIM_s}$     &  \textcolor{blue}{\textbf{0.393}} & \textcolor{green}{\textbf{0.322}} & \textcolor{green}{\textbf{0.210}}    & \textcolor{green}{\textbf{0.145}}  & \textcolor{green}{\textbf{0.150}}  \\ \hline 
    \end{tabular}
    }
    \label{tab:quantitative-results-test}
\end{table}

\subsection{Qualitative Results}\label{sec:qualitative-results}

\begin{figure*}
    \centering
    \begin{tabular}{c c c c}
         \includegraphics[width=0.20\linewidth]{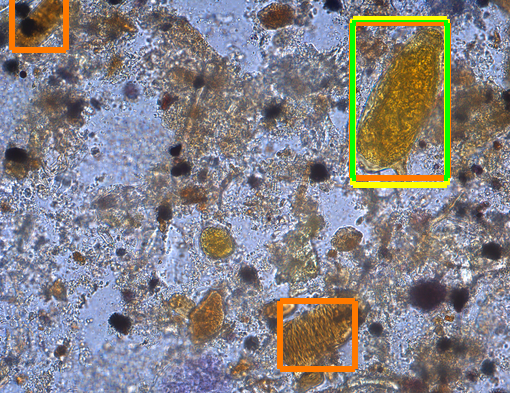} &
         \includegraphics[width=0.20\linewidth]{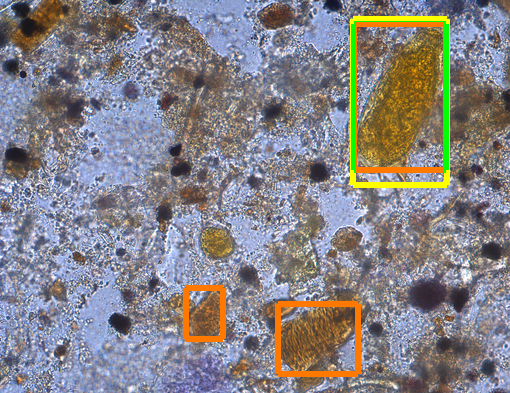} &
         \includegraphics[width=0.20\linewidth]{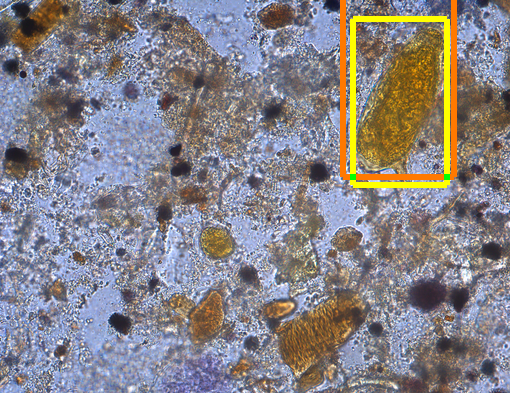} &
         \includegraphics[width=0.20\linewidth]{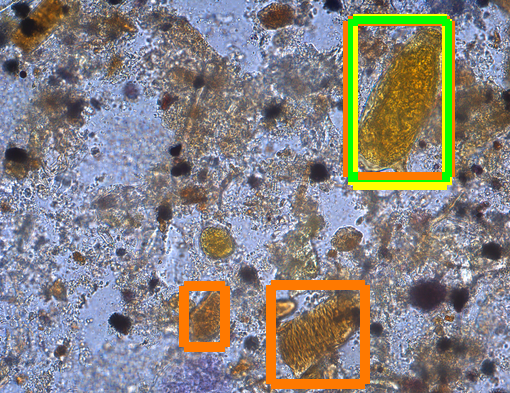} \\
         \includegraphics[width=0.20\linewidth]{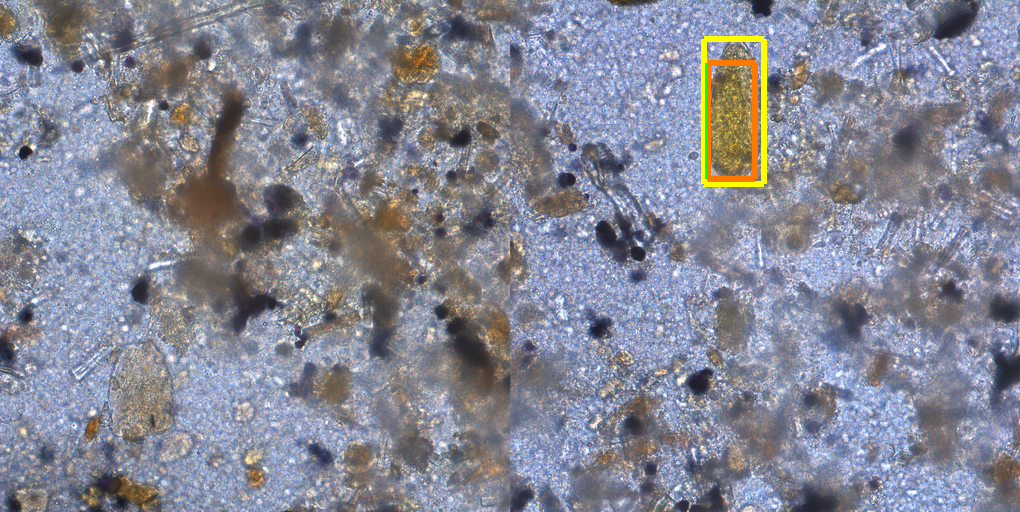} &
         \includegraphics[width=0.20\linewidth]{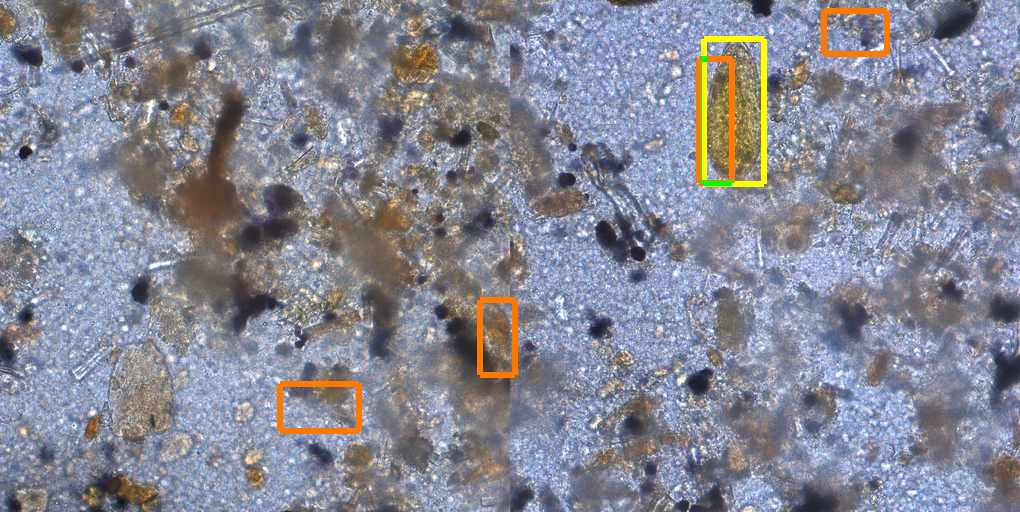} &
         \includegraphics[width=0.20\linewidth]{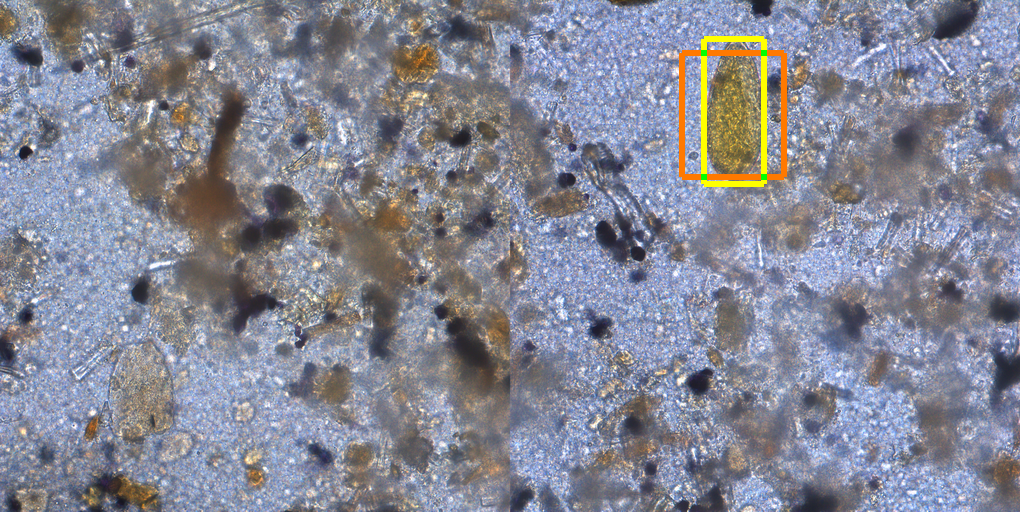} &
         \includegraphics[width=0.20\linewidth]{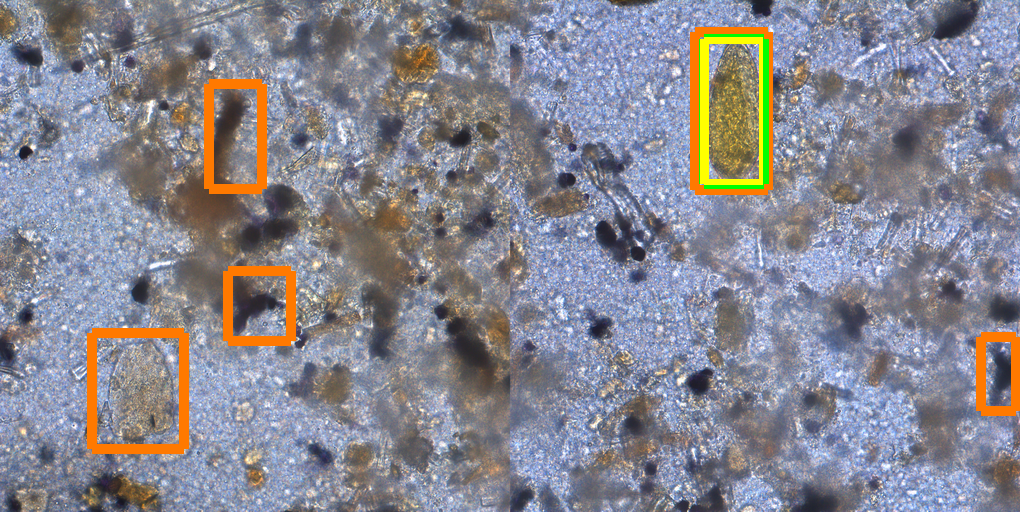} \\
         \includegraphics[width=0.20\linewidth]{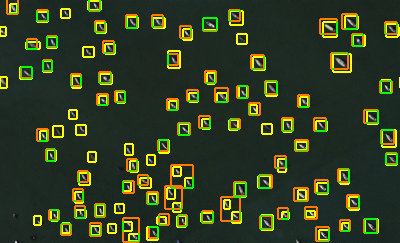} &
         \includegraphics[width=0.20\linewidth]{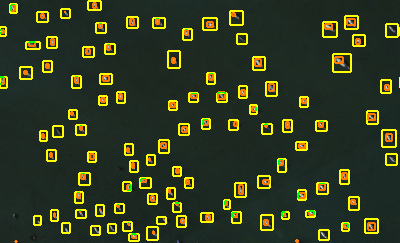} &
         \includegraphics[width=0.20\linewidth]{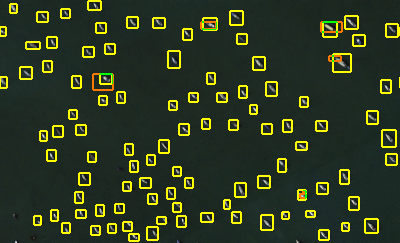} &
         \includegraphics[width=0.20\linewidth]{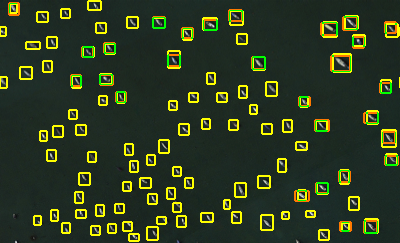} \\
         \includegraphics[width=0.20\linewidth]{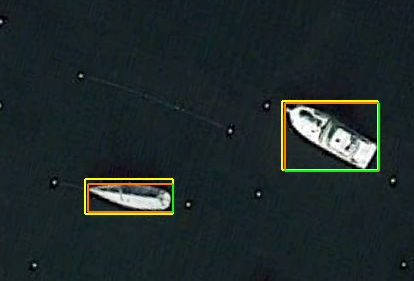} &
         \includegraphics[width=0.20\linewidth]{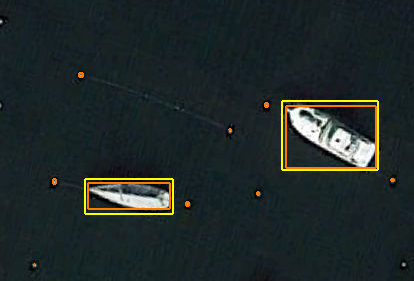} &
         \includegraphics[width=0.20\linewidth]{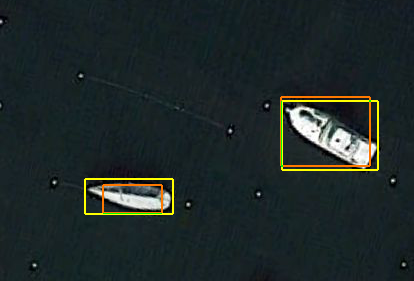} &
         \includegraphics[width=0.20\linewidth]{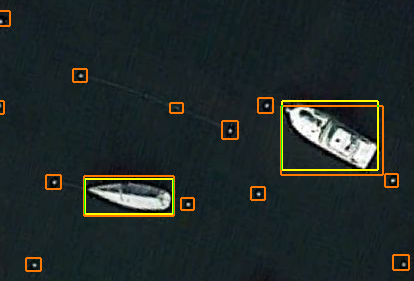} \\
         (a) \textbf{Adaptive-}$\mathbf{FLIM}$ & (b) U²Net & (c) Self-Reformer & (d) DETReg\\
     \end{tabular}
    \caption{Successful detection examples. The ground-truth boxes are in yellow, the predictions in orange, and their intersection in green.
    }
    \label{fig:quali-results}
\end{figure*}

\begin{figure*}
    \centering
    \begin{tabular}{c c c c}
         \includegraphics[width=0.22\linewidth]{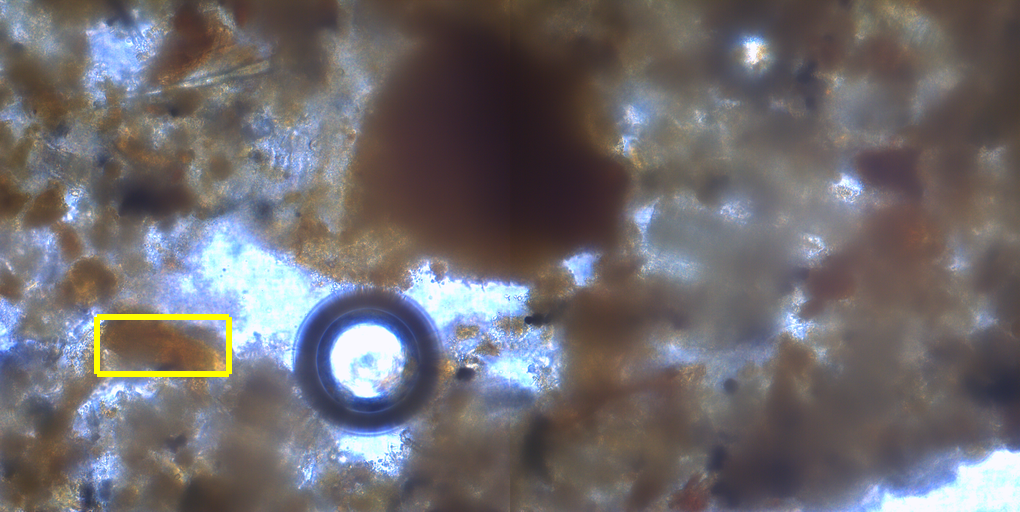} &
         \includegraphics[width=0.22\linewidth]{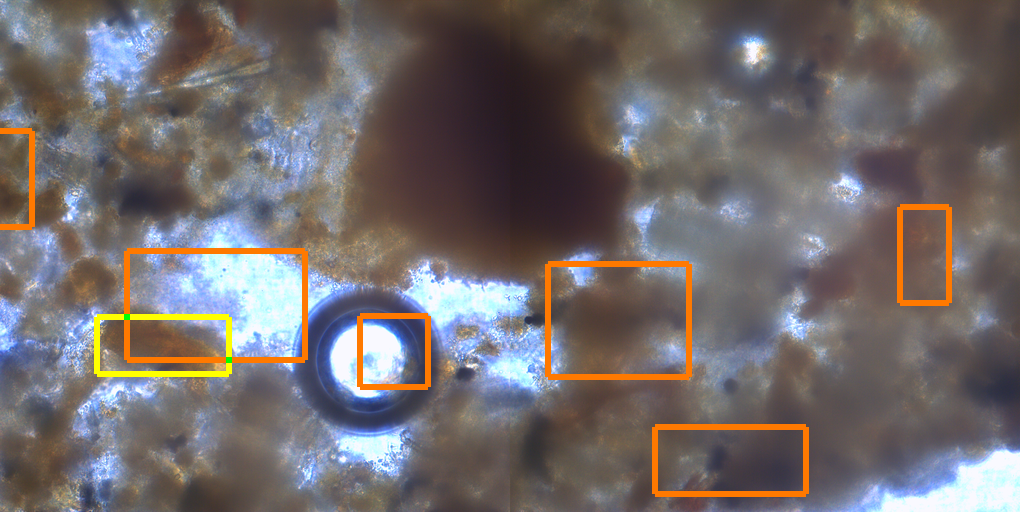} &
         \includegraphics[width=0.22\linewidth]{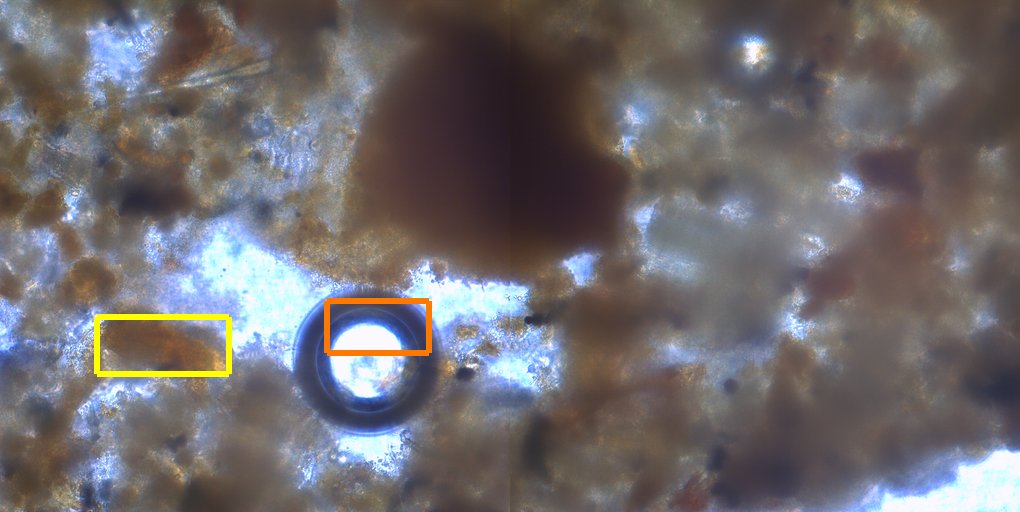} &
         \includegraphics[width=0.22\linewidth]{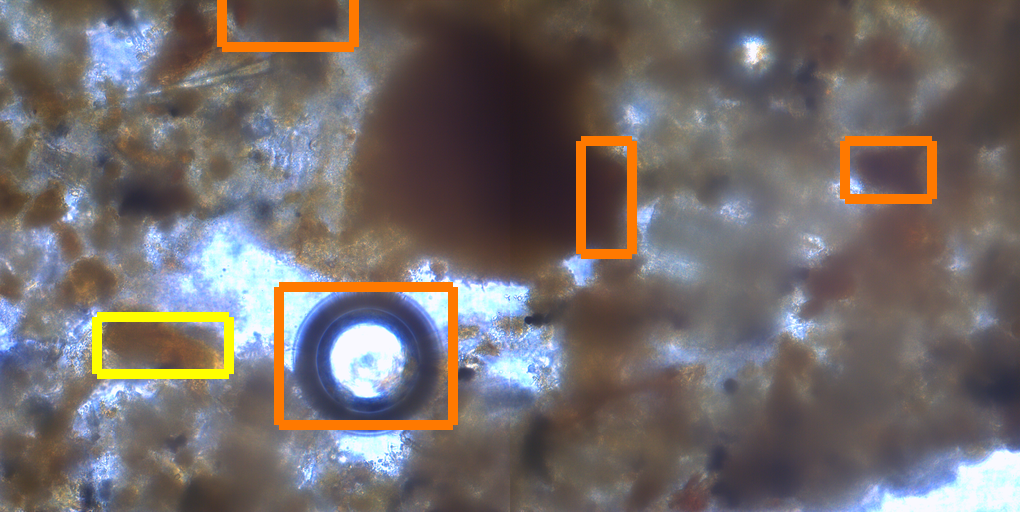} \\
         \includegraphics[width=0.22\linewidth]{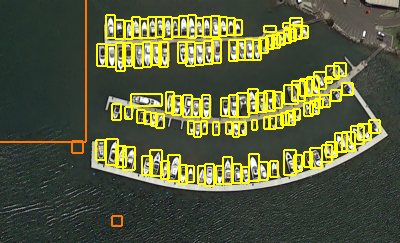} &
         \includegraphics[width=0.22\linewidth]{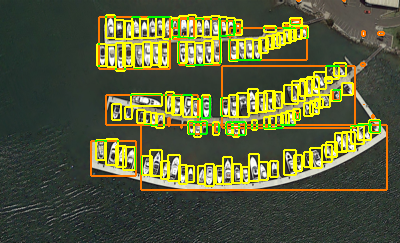} &
         \includegraphics[width=0.22\linewidth]{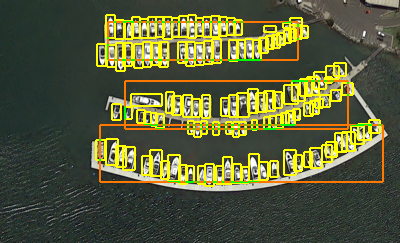} &
         \includegraphics[width=0.22\linewidth]{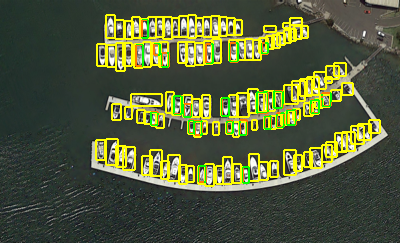} \\         
         \includegraphics[width=0.22\linewidth]{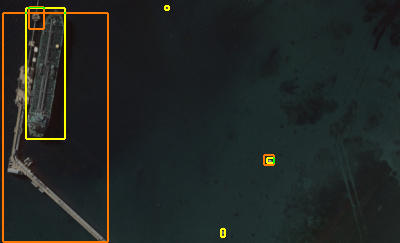} &
         \includegraphics[width=0.22\linewidth]{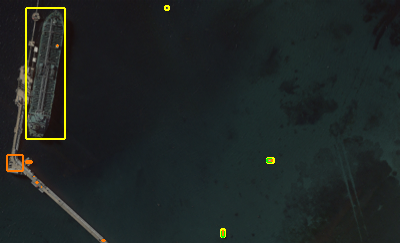} &
         \includegraphics[width=0.22\linewidth]{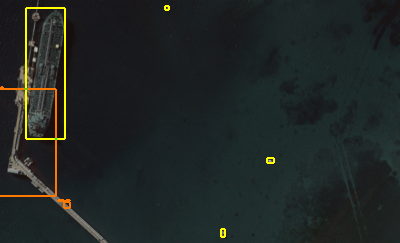} &
         \includegraphics[width=0.22\linewidth]{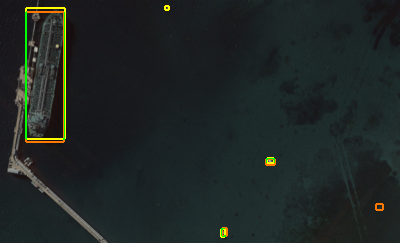} \\
         (a) \textbf{Adaptive-}$\mathbf{FLIM}$ & (b) U²Net & (c) Self-Reformer & (d) DETReg\\
     \end{tabular}
    \caption{Failure cases. The ground-truth boxes are in yellow, the predictions in orange, and their intersection in green.
    }
    \label{fig:quali-results-fails}
\end{figure*}

We visually compare our solution and the other methods. In Figure~\ref{fig:quali-results}, we show examples of images where our solution detected the correct component, while Figure~\ref{fig:quali-results-fails} depicts cases in which the object of interest was undetected.
    
   Starting from the success cases (Figure \ref{fig:quali-results}), in the parasite dataset, \textbf{Adaptive-}$\mathbf{FLIM}$ provides a small number of false positives and bounding boxes with good IoU. As discussed in Section \ref{subsec:quantitative}, U²Net has closer results to \textbf{Adaptive-}$\mathbf{FLIM}$, but at the cost of more false positives. Self-Reformer, on the other hand, can detect most objects with a low rate of false positives but with poor bounding boxes that often are below the IoU thresholds. DETReg estimates high-IoU boxes for the objects of interest but has the highest false positives of all methods. Regarding the ship dataset, FLIM has the largest recall, detecting small objects with good $IoU$ scores compared to the other methods that either missed (DETReg, Self-reformer) or poorly detected (U²Net). It could also distinguish small non-ship objects, decreasing the number of false positives (last row). 

   Regarding the failure cases (Figure \ref{fig:quali-results-fails}), because of how our adaptation heuristic works, it would be hard to identify foreground kernels in images with cluttered backgrounds containing mostly structures similar to the object of interest, especially if they are connected to the objects. However, the other methods also fail to detect the object. Additionally, for the ships dataset, when many objects are very close to each other, the saliency estimators estimate one bounding box for each connected component, missing the individual objects. Even though DETReg better handles connected objects, it also missed several.

\begin{table}[htb]
 \centering
    \caption{Comparison among the number of parameters of different models. The deep models presented are used in our comparisons, and the lightweight models are standard backbones used for object detection. Models with (*) indicate no manual kernel selection}
    \resizebox{.6\linewidth}{!}{%
    \begin{tabular}{l | c }
\rowcolor{gray!25}
    \hline
    \textbf{Deep Models}            & \#Parameters  \\ \hline
    DETReg                          & 39.847.265\\ \hline 
    U²Net                           & 44.009.869\\ \hline
    Self-Reformer                   & 91.585.457\\ \hline  \hline
\rowcolor{gray!25} 
    \textbf{Lightweight Models}     & \#Parameters \\ \hline
    MobileNetv2                     & 3.504.872 \\ \hline 
    SqueezeNet                      & 1.248.424 \\ \hline  \hline 
\rowcolor{gray!25}    
    \textbf{Flyweight Models}       & \#Parameters \\ \hline 
    Adaptive-$FLIM_s$*              & 93,350    \\ \hline 
    \textcolor{blue}{\textbf{Adaptive-}$\mathbf{FLIM_s}$} & \textcolor{blue}{\textbf{7,931}}      \\ \hline
    Adaptive-$FLIM_p$*              & 53,333    \\ \hline 
    \textcolor{blue}{\textbf{Adaptive-}$\mathbf{FLIM_p}$} & \textcolor{blue}{\textbf{2,296}}      \\ \hline
    
    \end{tabular}
    }
    \label{tab:model-sizes}
    % \vspace{-0.5cm}
\end{table}

\section{Conclusion}\label{sec:conclusion}
    We presented a user-in-the-learning-loop methodology, named \textbf{Adaptive-FLIM}, to improve FLIM-based encoders and build CNNs layer by layer under the user's control in the number of layers and kernels per layer. Kernel selection is based on visual analysis of activations and object saliency maps. We introduced a single-layer adaptive decoder to create those saliency maps at the output of any layer. The final CNN consists of a FLIM-based encoder followed by the adaptive decoder. We showed that \textbf{Adaptive-FLIM} could create flyweight networks for object detection, outperforming or presenting competitive results to SOTA approaches in two datasets using five metrics. In both cases,  \textbf{Adaptive-FLIM}  required only five images to train from scratch without segmentation masks while being thousands of times smaller than the baselines. This result indicates that the proposed methodology has great potential to provide embedded solutions.  
    
    We intend to explore multi-scale feature extraction and extend our approach to multi-class object detection.

\section*{Acknowledgments}
This work was supported by ImmunoCamp, CAPES (88887.191730/2018-00), CNPq (303808/2018-7, 407242/2021-0, 306573/2022-9) and FAPESP (2014/12236-1).

%\newpage
\bibliographystyle{IEEEtran}
\bibliography{main}
\vskip -2\baselineskip plus -1fil
\begin{IEEEbiographynophoto}{Leonardo de Melo Joao}
Leonardo de Melo Joao received a M.Sc. degree in computer science from the University of Campinas, in 2021, and a B.Sc. degree in computer science from the Pontifical Catholic University of Minas Gerais in 2018, both in Brazil. He is currently pursuing the Ph.D. degree in computer science from the University of Campinas, Campinas, and his research interests include image processing, and machine learning.
\end{IEEEbiographynophoto}
\vskip -2\baselineskip plus -1fil
\begin{IEEEbiographynophoto}{Azael de Melo e Sousa}
Azael de Melo e Sousa received a M.Sc. degree in computer science from the University of Campinas, in 2017, and a B.Sc. degree in computer science from the São Paulo State University in 2015. He is currently a Ph.D. student in computer science on the University of Campinas, whose research focus on image processing and machine learning for medical image analysis.
\end{IEEEbiographynophoto}
\vskip -2\baselineskip plus -1fil
\begin{IEEEbiographynophoto}{Bianca Martins dos Santos}
Bianca Martins dos Santos received the B. Sc. degree in Biomedicine from Faculdade Integrada Metropolitana de Campinas (METROCAMP) (2009), Master's degree in Animal Biology in Parasitology (2015) and a PhD in Sciences (2021) both from the State University of Campinas (UNICAMP). She has experience in the area of Parasitology, with an emphasis on Diagnosis of Intestinal Parasites. She is currently a researcher at the Laboratory of Image Data Science (LIDS) at UNICAMP.
\end{IEEEbiographynophoto}
\vskip -2\baselineskip plus -1fil
\begin{IEEEbiographynophoto}{Silvio Jamil Ferzoli Guimarães}
Silvio Jammil Ferzolli Guimarães holds a Ph.D. in Computer Science from the Federal University of Minas Gerais (2003) and a joint Ph.D. in Informatique - Universite de Marne La Vallee (2003). He is currently Professor at the Pontifical Catholic University of Minas Gerais (PUC Minas) and an Associate Researcher at ESIEE / Paris and fellowship of research productivity PQ 2. He has experience in Computer Science, working mainly in digital video analysis and processing, mathematical morphology, graph-based image and video processing.
\end{IEEEbiographynophoto}
\vskip -2\baselineskip plus -1fil
\begin{IEEEbiographynophoto}{Jancarlo Ferreira Gomes}
Graduated in Agronomic Engineering from the State University of Northern Paraná, Master and Doctorate in Parasitology from the Institute of Biology at the University of Campinas, and Post-Doctorate in Public Health at the University of Campinas. He is currently full professor and colaborator at the Faculty of Medical Sciences and Researcher at the Institute of Computing at the University of Campinas, with experience in the areas of Human and Animal Parasitology and Computer Science.
\end{IEEEbiographynophoto}
\vskip -2\baselineskip plus -1fil
\begin{IEEEbiographynophoto}{Ewa Kijak}
received a Ph.D. in Computer Science from the University of Rennes, France (2003). She is currently associate professor at Université de Rennes, within the IRISA laboratory. Her research interests cover different topics of computer vision as machine learning, image analysis, description and indexing, as well as multimodal applications.

\end{IEEEbiographynophoto}
\vskip -2\baselineskip plus -1fil
\begin{IEEEbiographynophoto}{Alexandre Xavier Falcao}
received a B.Sc. (1988) in Electrical Engineering from the Federal University of Pernambuco, Brazil, an M.Sc. (1993), and a Doctorate (1997) in Electrical Engineering from the University of Campinas, Brazil. He is a full professor at the University of Campinas and holds a CNPq research productivity fellowship PQ 1A. His research interests are graph-based image processing, multidimensional data visualization, machine learning, image analysis, and their applications in the Sciences and Engineering. 
\end{IEEEbiographynophoto}

\end{document}